%% file: gdcn.tex
\documentclass[sigconf]{acmart}
\usepackage{multirow}
\usepackage{makecell}
\usepackage{subfig}
\usepackage{bbding}
\usepackage{enumitem}
\usepackage{colortbl}
\usepackage{diagbox}
\usepackage{balance}

\copyrightyear{2023}
\acmYear{2023}
\setcopyright{acmlicensed}
\acmConference[CIKM '23] {Proceedings of the 32nd ACM International Conference on Information and Knowledge Management}{October 21--25, 2023}{Birmingham, United Kingdom.}
\acmBooktitle{Proceedings of the 32nd ACM International Conference on Information and Knowledge Management (CIKM '23), October 21--25, 2023, Birmingham, United Kingdom}
\acmPrice{15.00}
\acmISBN{979-8-4007-0124-5/23/10}
\acmDOI{10.1145/3583780.3615089}

\begin{document}

\title{Towards Deeper, Lighter and Interpretable Cross Network for CTR Prediction}

\author{Fangye Wang}
\additionalaffiliation{
Shanghai Key Laboratory of Data Science, Shanghai Institute of Intelligent Electronics \& Systems, China.
}
\orcid{0000-0001-7216-1688}
\affiliation{
{
\normalsize
  \institution{School of Computer Science}
  \institution{Fudan University
  \city{Shanghai}
  \country{China}}
}
}
\email{fywang18@fudan.edu.cn}

\author{Hansu Gu}
\orcid{0000-0002-1426-3210}
\affiliation{
{
\normalsize
  \city{Seattle}
  \country{United States}
}
}
\email{hansug@acm.org}

\author{Dongsheng Li}
\orcid{0000-0003-3103-8442}
\affiliation{
{
\normalsize
  \institution{Microsoft Research Asia}
  \city{Shanghai}
  \country{China}
 }
}
\email{dongsli@microsoft.com}

\author{Tun Lu}
\authornotemark[1]
\authornote{Corresponding author.}
\orcid{0000-0002-6633-4826}
\affiliation{
\normalsize
  \institution{School of Computer Science}
  \institution{Fudan University 
  \city{Shanghai} 
  \country{China}}
 }
\email{lutun@fudan.edu.cn}

\author{Peng Zhang}
\authornotemark[1]
\orcid{0000-0002-9109-4625}
\affiliation{
 \normalsize
  \institution{School of Computer Science}
  \institution{Fudan University
  \city{Shanghai}
  \country{China}
  }
 }
\email{zhangpeng_@fudan.edu.cn}

\author{Ning Gu}
\authornotemark[1]
\orcid{0000-0002-2915-974X}
\affiliation{
{
  \normalsize
  \institution{School of Computer Science}
  \institution{Fudan University
  \city{Shanghai}
  \country{China}}
 }
}
\email{ninggu@fudan.edu.cn}

\renewcommand{\shortauthors}{Fangye Wang et al.}

\begin{abstract}

Click Through Rate (CTR) prediction plays an essential role in recommender systems and online advertising. It is crucial to effectively model feature interactions to improve the prediction performance of CTR models. However, existing methods face three significant challenges. First, while most methods can automatically capture high-order feature interactions, their performance tends to diminish as the order of feature interactions increases. Second, existing methods lack the ability to provide convincing interpretations of the prediction results, especially for high-order feature interactions, which limits the trustworthiness of their predictions. Third, many methods suffer from the presence of redundant parameters, particularly in the embedding layer. This paper proposes a novel method called Gated Deep Cross Network (GDCN) and a Field-level Dimension Optimization (FDO) approach to address these challenges. As the core structure of GDCN, Gated Cross Network (GCN) captures explicit high-order feature interactions and dynamically filters important interactions with an information gate in each order. Additionally, we use the FDO approach to learn condensed dimensions for each field based on their importance. Comprehensive experiments on five datasets demonstrate the effectiveness, superiority and interpretability of GDCN. Moreover, we verify the effectiveness of FDO in learning various dimensions and reducing model parameters. The code is available on \url{https://github.com/anonctr/GDCN}.


\end{abstract}



\maketitle
\section{Introduction}

Click-through Rate (CTR) prediction is an important component of recommender systems and online advertising~\cite{cheng2016wide, covington2016youtobednn}. It aims to estimate the probability of a user clicking on a recommended item or an advertisement on a web page. Accurate CTR prediction can bring significant revenue gains and also improved user satisfaction~\cite{cheng2016wide, wang2021masknet, zhou2019dien}. Most methods typically consist of three layers~\cite{zhang2021deep, zhu2022bars, wang2022enhancing}: feature embedding, feature interaction, and prediction. To improve the accuracy of CTR prediction, many methods have been proposed that focus on designing effective feature interaction architectures. However, previous works such as Logistic Regression (LR)\cite{richardson2007predicting}, and FM-based methods~\cite{rendle2012factorization, blondel2016higher, juan2016field} can only model low- or fixed-order feature interactions. As web-scale recommendation systems have become more complex, there is a growing demand for methods to capture high-order feature interactions. Therefore, more recent methods~\cite{lian2018xdeepfm, wang2021dcnm, wang2017deep, song2019autoint, cheng2020adaptive, yu2020deepim, guo2019oenn} enable joint modeling of both explicit and implicit high-order feature interactions and achieve significant performance improvements. While these methods have made great progress, they still have three major limitations.

First, the effectiveness of these methods tends to decrease as the order of feature interactions increases. In general, the maximum degree of interactions that can be captured is determined by the depth of feature interactions. As the interaction layers go deeper, the number of interactions increases exponentially, which enables the model to generate more high-order interactions. However, not all interactions are helpful, which also brings in many unnecessary interactions, resulting in decreased performance and increased computational complexity. Many existing state-of-the-art (SOTA) works~\cite{covington2016youtobednn, lian2018xdeepfm, cheng2020adaptive, blondel2016higher, song2019autoint, wang2017deep, wang2021dcnm, guo2017deepfm, qu2018product, zhao2021fint, he2017neural, lang2021architecture} have confirmed through hyper-parameter analysis that their performance deteriorates when the interaction order exceeds a certain depth, usually three orders. Therefore, it is crucial to make improvements and ensure that the high-order interactions have a positive impact rather than introducing more noise and lead to sub-optimal performance.

Second, the lack of interpretability in existing methods limits the trustworthiness of their predictions and recommendations. Most methods~\cite{shan2016deepcrossing, qu2018product, he2017neural, guo2017deepfm, zhang2016fnn} suffer from low interpretability due to the implicit feature interactions through DNNs or the assignment of equal weights to all feature interactions~\cite{chen2021dcap, song2019autoint}. Although a few methods~\cite{chen2021dcap, song2019autoint, li2020interpretable} have attempted to provide explanations through attention scores learned by the Self-attention mechanism~\cite{vaswani2017attention}, this approach tends to fuse all the features information, making it difficult to distinguish which interactions are essential, especially for high-order crosses. Hence, it is vital to develop methods that can provide a persuasive interpretation from both the model and instance perspectives, enabling more reliable and trustworthy results.

Third, most existing models contain massive redundant parameters, particularly in the embedding layer. Many methods~\cite{rendle2012factorization, chen2021dcap, song2019autoint, lian2018xdeepfm, zhao2021fint, mao2023finalmlp, yan2023dynint} rely on feature-wise interaction structures, which assume equal embedding dimensions for all fields. However, some fields only require a relatively short dimension considering their information capacity. Consequently, these models produce massive redundant parameters in the embedding layer. But directly reducing the embedding dimension leads to a decrease in model performance~\cite{sun2020generic, song2019autoint, he2017ncf, he2017neural}. Meanwhile, most methods~\cite{tian2023directed, wang2021dcnm, yan2023dynint, zhang2022fibinet++} only focus on reducing non-embedding parameters, and the impact on overall parameter reduction is not significant compared to the embedding parameters. Although DCN~\cite{wang2017deep} and DCN-V2~\cite{wang2021dcnm} assign varying dimensions to each field using a rule-of-thumb formula which computes dimension only based on feature numbers, they overlook the importance of each field and often fail to reduce model parameters. Hence, we aim to assign field-specific and condensed dimensions to each field, considering their inherent importance and effectively reducing the embedding parameters.

This paper presents a model called Gated Deep Cross Network (GDCN) and a approach called Field-level Dimension Optimization (FDO) to address the above-mentioned limitations. Building upon the elegant and efficient design of DCN-V2~\cite{wang2021dcnm}, GDCN further offers improved performance in both low-order and high-order interactions, and also shows great interpretability at both model and instance perspectives. GDCN models explicit feature interactions through a proposed Gated Cross Network (GCN) and then integrates with a DNN to learn implicit feature interactions. GCN consists of two core components: feature crossing and information gate. The feature crossing component captures explicit interactions within a bounded degree, while the information gate selectively amplifies important cross features with high importance and mitigates the influence of unimportant ones at each cross order. Additionally, considering their respective importance, the FDO approach can allocate condensed and independent dimensions to each field.

The core contributions of this work are summarized as follows:
\begin{itemize}[leftmargin=0.4cm]
 \item We introduce a novel method GDCN to learn both explicit and implicit feature interactions through GCN and DNN. GCN designs an information gate to dynamically filter the next-order cross features and effectively control the information flow. Compared to existing methods, GDCN demonstrates improved performance and stability in capturing deeper high-order interactions.
 \item We develop the FDO approach to assign condensed dimensions to each field, considering their inherent importance. By employing FDO, GCN achieves comparable performance with only 23\% of the original model parameters and outperforms existing SOTA models with smaller model size and faster training speed. 
 \item Comprehensive experiments show great effectiveness and generalization of GDCN on five datasets. Moreover, our methods provide remarkable interpretability at the model and instance levels, enhancing our understanding of the model predictions.
\end{itemize}

\section{Related work}
\subsection{CTR Prediction}
Modeling informative feature interactions has been widely studied in the field to improve the performance of CTR models. Traditional methods such as LR~\cite{richardson2007predicting}, and FM-based methods~\cite{rendle2012factorization, blondel2016higher, juan2016field} model low-order feature interactions. Recent deep learning-based methods have made significant progress by capturing high-order feature interactions. These works can be divided into two categories: stacked and parallel models, based on how they integrate explicit and implicit interaction networks~\cite{chen2021enhancing, wang2022enhancing}.

\textbf{Stacked Structure.} Stacked models first employ an interaction network to capture explicit interactions on top of the embedding layer and then use a network such as DNN to further model implicit interactions. Representative explicit structures include inner product and outer product (e.g., PNN~\cite{qu2018product}, ONN~\cite{yang2020operation}), Hadamard product (e.g., FM~\cite{rendle2012factorization}, FFM~\cite{pan2018field}), cross network (e.g., CN~\cite{wang2017deep}, CN-V2~\cite{wang2021dcnm}, XCrossNet~\cite{yu2021xcrossnet}), Bi-Interaction (e.g., NFM~\cite{he2017neural}), attention operation (e.g., AFM~\cite{xiao2017attentional}, AutoInt~\cite{song2019autoint}, DCAP~\cite{chen2021dcap}, and DIEN~\cite{zhou2019dien}). After capturing explicit interactions, DNN is used to model deeper implicit interactions based on the explicit interactions output.

\textbf{Parallel Structure.} Parallel models jointly capture explicit and implicit interaction information with two parallel networks. Representative parallel models include WDL~\cite{cheng2016wide}, DeepFM~\cite{guo2017deepfm}, DCN~\cite{wang2017deep}, DCN-V2~\cite{wang2021dcnm}, xDeepFM~\cite{lian2018xdeepfm} and AFN+~\cite{cheng2020adaptive}. Among these models, DNN is the most commonly used and efficient network to capture implicit interactions over the embedding layer. The main difference among these models lies in how they model explicit interactions. WDL adopts LR as the wide part to enhance memorization ability. DeepFM utilizes FM~\cite{rendle2012factorization} operation to capture pair-wise interactions adaptively. DCN and DCN-V2 propose two kinds of cross networks(i.e., CN and CN-V2) to extract bounded-degree feature interactions automatically. xDeepFM designs a CIN structure to capture complex feature interactions of bounded orders. Some other models jointly train three or more parallel networks to achieve better performance, e.g., FED~\cite{zhao2020fed}, NON~\cite{luo2020network}, MaskNet~\cite{wang2021masknet}. 

Despite these models can capture high-order feature crosses, they experience a decline in performance as the cross layers go higher. And they lack interpretability in identifying important crosses at both model and instance levels. Moreover, most models assign equal dimensions for all fields, leading to massive redundancy parameters. This paper aims to address these issues with our proposed methods.

\subsection{Gating Mechanism in CTR Prediction}
The gating mechanism has been widely adopted in various well-known methods, such as LSTM~\cite{hochreiter1997lstm}, GRU~\cite{cho2014gru}, MMoE~\cite{ma2018mmoe}. Generally, gates enable the selection of essential features or control information flow by assigning different levels of importance to different features or sub-networks. Specifically, Multi-gate Mixture-of-Experts (MMoE)~\cite{ma2018mmoe} adopts several gating networks to weight the importance of several task-specific objectives. The idea of MMoE is applied in DCN-V2~\cite{wang2021dcnm} and DynInt~\cite{yan2023dynint} to reduce non-embedding parameters for better cost-efficiency through a mixture of experts or weight matrix decomposition. IFM~\cite{yu2019input}, DIFM~\cite{lu2020dual}, and FiBiNet~\cite{huang2019fibinet} propose different weight learning networks to recalibrate the importance of feature embeddings. Other works~\cite{wang2021contextnet, wang2022mcrf, wang2021masknet, huang2020gatenet, fei2021gemnn} propose different structures to select salient information from both feature embedding and intermediate representations in bit-level. In this paper, GDCN designs the information gate to identify important cross features in each cross layer, particularly for higher-order cross features. This enables GDCN to mitigate the noise introduced by exponential high-order interactions and provides dynamic interpretability to identify critical interactions for each instance.

\section{Proposed Architecture}
Inspired by DCN-V2~\cite{wang2021dcnm}, we develop GDCN, which consists of an embedding layer, gated cross network (GCN) and deep network (DNN). The embedding layer transforms the high-dimensional sparse input into low-dimensional dense representations. The GCN is designed to capture explicit feature interactions, with an information gate to identify the important cross features. Then, a DNN is integrated to model implicit feature crosses. In essence, GDCN is a generalization of DCN-V2, inheriting the excellent expressiveness of DCN-V2 with a simple and elegant formula for easy deployment. However, GDCN introduces a key difference by incorporating information gates, which adaptively filter the cross features in each order instead of uniformly aggregating all features. This enables GDCN to truly utilize deeper high-order cross-information without experiencing performance degradation and empowers GDCN with dynamic interpretability for each instance. The architecture of GDCN is depicted in Figure \ref{fig:gdcn}, showing two structures that combine the GCN and DNN networks: (a) GDCN-S and (b) GDCN-P.

\subsection{Embedding Layer}
In the CTR prediction task, the input features (e.g., categorical and numerical features) are typically high-dimensional and sparse~\cite{meng2021autopi, song2019autoint, zhao2021non}. Input instances are usually multi-field tabular
data records~\cite {wang2023cl4ctr, guo2022miss, zhao2021non}, which contain $F$ different fields and $T$ features. Each instance is represented by a field-aware one-hot vector~\cite{song2019autoint, wang2022enhancing, meng2021autopi}. The embedding layer transforms the sparse high-dimensional features into a dense low-dimensional embedding matrix $\mathbf{E}=[\mathbf{e_1};...;\mathbf{e_F}]$. Most CTR models~\cite{naumov2019deep, qu2018product, huang2019fibinet, song2019autoint, meng2021autopi, guo2017deepfm} require embedding dimension to be the same to accommodate specific interaction operations. However, GDCN allows arbitrary embedding dimensions, and the output of the embedding layer is represented by the concatenated vector $\mathbf{c_0}=[\mathbf{e_1}\parallel ... \parallel \mathbf{e_F}] \in \mathbb{R} ^ {D}$.
 
\subsection{Gated Cross Network (GCN)}
\begin{figure}[t]
    \setlength{\abovecaptionskip}{0.2cm}
    \setlength{\belowcaptionskip}{-0.2cm}
    \centering
    \includegraphics[width=0.42\textwidth]{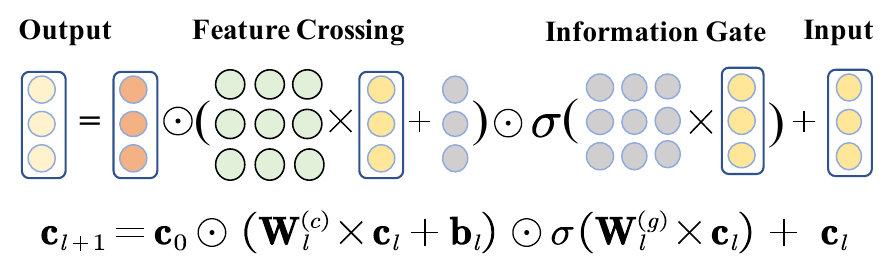}
    \caption{Visualization of the gated cross layer.$\odot$ is element-wise/Hadamard product, and $\times$ is matrix multiplication.}
    \label{fig:gcn}
\end{figure}

As the core structure of GDCN, the GCN aims to model explicit bounded-degree feature crosses with information gate. The $(l+1)^{th}$ gated cross layer of the GCN is represented by:
\begin{small}
\begin{equation}
\label{equ:gcn}
\mathbf{c}_{l+1}=\underset{Feature\,\,Crossing}{\underbrace{\mathbf{c}_0\odot \left( \mathbf{W}_{l}^{\left( c \right)} \times \mathbf{c}_l+\mathbf{b}_l \right) }}\odot \underset{Information\,\,Gate}{\underbrace{\sigma \left( \mathbf{W}_{l}^{\left( g \right)} \times \mathbf{c}_l \right) }}+\mathbf{c}_l,\\
\end{equation}
\end{small}
where $\mathbf{c}_0$ is the base input from the embedding layer, which contains the 1st-order features; $\mathbf{c}_{l} \in \mathbb{R}^D $ are the output features from the previous $l^{th}$ gated cross layer and used as input to the current $(l+1)^{th}$ layer, and $\mathbf{c}_{l+1} \in \mathbb{R}^D $ are the output; $\mathbf{W}_{l}^{(c)}$, $\mathbf{W}_{l}^{(g)} \in \mathbb{R}^{D\times D}$ and $\mathbf{b}_{l} \in \mathbb{R}^{D}$ are the two learnable matrices and the bias vector, respectively. Figure \ref{fig:gcn} visualizes the process of the gated cross layer.

In each gated cross layer, there are two core components: the \textit{feature crossing} and the \textit{information gate} as shown in Equ.(\ref{equ:gcn}) and Figure \ref{fig:gcn}. The \textit{feature crossing} component calculates the interaction between the 1st-order feature $\mathbf{c}_0$ and the $(l+1)^{th}$ order features $\mathbf{c}_{l}$ in bit-level. It then outputs the next polynomial order interaction that contains all $(l+2)^{th}$ order cross features. The matrix $\mathbf{W}_l^{(c)}$, known as the cross matrix, indicates the inherent importance among various fields in the $(l+1)^{th}$ order. However, not all $(l+2)^{th}$ order features have a positive impact on the prediction. As cross depth increases, the cross features exhibit exponential growth, introducing cross noise that can lead to sub-optimal performance. To address this, the \textit{information gate} component is introduced to act as a soft gate that adaptively learns the importance of $(l+2)^{th}$ order features. The gate values are obtained by applying the sigmoid function $\sigma(\cdot)$ to the result of matrix multiplication between the gate matrix $\mathbf{W}_l^{(g)}$ and the input $\mathbf{c}_{l}$. They are then element-wise multiplied with the output of the feature crossing component that contains unselected $(l+2)^{th}$ order cross features. This process amplifies important features and mitigates the impact of unimportant features. As the number of cross layers increases, the information gate at each cross layer filters the next-order cross features and effectively controls the information flow. Finally, the ultimate output cross vector $\mathbf{c}_{l+1}$ is generated by adding the input $\mathbf{c}_{l}$ to the result of the feature crossing and information gate, thus containing all the feature interactions from the 1st order to the $(l+2)^{th}$ order. 


\begin{figure}[t]
    \setlength{\abovecaptionskip}{0.2cm}
    \setlength{\belowcaptionskip}{-0.2cm}
    \centering
    \includegraphics[width=0.43\textwidth]{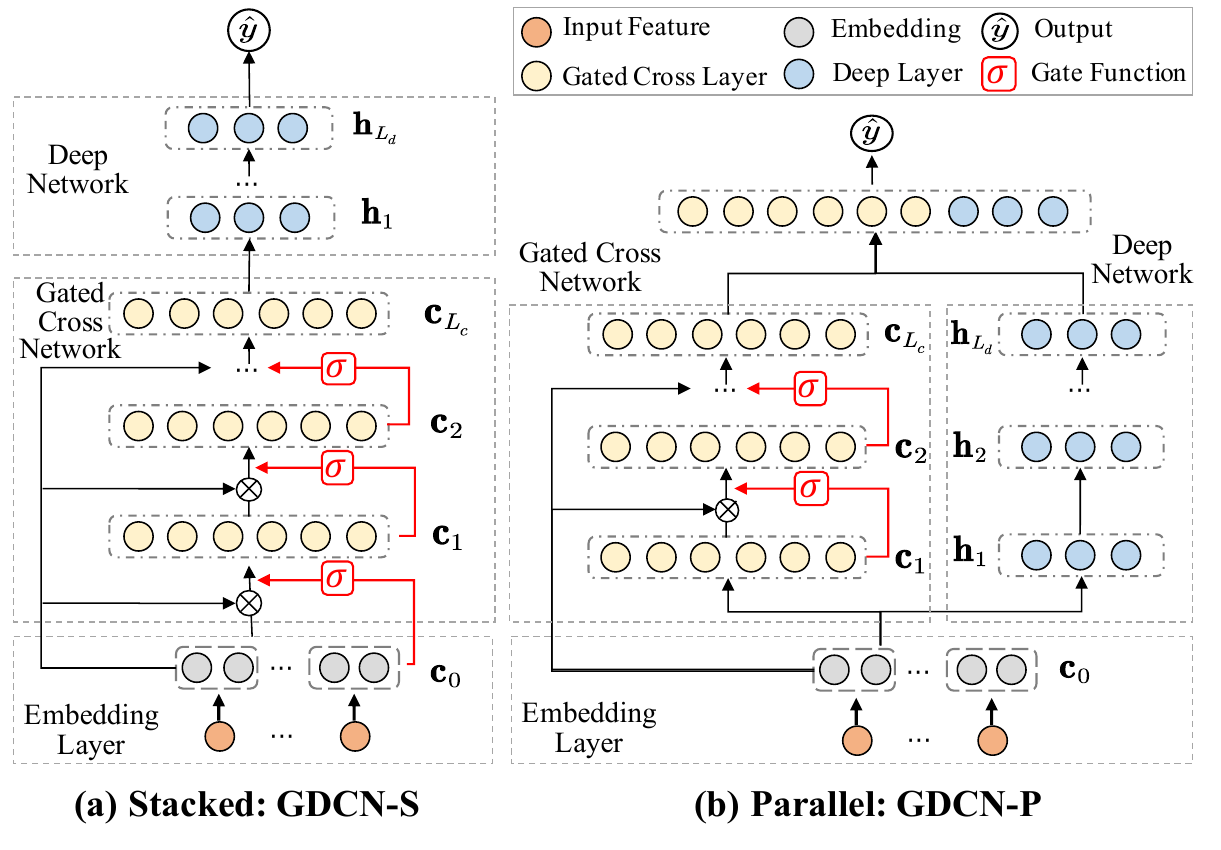}
    \caption{Architecture of the GDCN-S and GDCN-P. $\otimes$ is the cross operation(a.k.a, the gated cross layer) in Equation \ref{equ:gcn}.}.
    \label{fig:gdcn}
\end{figure}

\subsection{Deep Neural Network (DNN)}
The objective of the DNN is to model implicit feature interactions. Each deep layer of DNN is represented by $\mathbf{h}_{l+1} = f(\mathbf{W}_l\mathbf{h}_l+\mathbf{b}_l)$, where $\mathbf{W}_l \in \mathbb{R}^{n_{l+1}\times n_{l}}$, $\mathbf{b}_{l} \in \mathbb{R}^{n_{l+1}}$ are the weight matrix and bias vector in $l^{th}$ deep layer; and $\mathbf{h}_{l+1} \in \mathbb{R}^{n_{l+1}}$, $\mathbf{h}_{l} \in \mathbb{R}^{n_{l}}$ are the output and input. $f(\cdot)$ is the activation function, which is usually ReLU. 


\subsection{Combine GCN and DNN}
Existing works adopt two main structures to integrate explicit and implicit interaction information: stacked and parallel. We also have two versions of GDCN by combining the GCN and DNN. 

Figure  \ref{fig:gdcn}(a) shows the \textbf{stacked} structure: GDCN-S.
The embedding vector $\mathbf{c}_0$ is fed into the GCN and output $\mathbf{c}_{L_c}$, followed by a DNN to generate the final cross vector $\mathbf{c}_{final} = \mathbf{h}_{L_d}$. $L_c$ and $L_d$ are the depth of gated cross layer and deep network, respectively. 

Figure \ref{fig:gdcn}(b) shows the \textbf{parallel} structure: GDCN-P. Vector $\mathbf{c}_0$ is parallel fed into GCN and DNN. Their outputs (i.e., $\mathbf{c}_{L_c}$ and $\mathbf{h}_{L_d}$) are concatenated to obtain the final cross vector $\mathbf{c}_{final}=[\mathbf{c}_{L_c}\parallel\mathbf{h}_{L_d}]$.

\textbf{Prediction and Training.} Finally, we calculate the prediction click probability $\hat{y}_i$ by a standard logistic regression function: $\hat{y}_i = \sigma(\mathbf{w}_{logit}\mathbf{c}_{final})$, where $\mathbf{w}_{logit}$ is the weight vector and $\sigma(z) ={1}/(1+e^{-z})$. 

The loss function is the widely used binary cross-entropy loss (a.k.a, LogLoss)~\cite{wang2021dcnm, liu2019fgcnn, zhu2022bars}: 
\begin{align}
\textstyle
\mathcal{L}_{ctr} =-\frac{1}{N}\sum_{i=1}^N 
\left({y_i} 
    \log\left(\hat{y}_i \right) 
    +\left( 1-y_i \right) \log \left( 1-\hat{y}_i\right)
\right),
\end{align}
where $\hat{y}_i$ and $y_i$ are the predicted and the ground-truth click probability, respectively. $N$ is the number of all training instances.

\subsection{Relationship with DCN-V2} 

GDCN is a generalization of DCN-V2. When the information gate is omitted, or all gate values are set to 1, GDCN falls back to DCN-V2~\cite{wang2021dcnm}. In DCN-V2, the cross layer (i.e., CN-V2) treats all cross features equally and directly aggregates them to the next order without considering the varying importance of different cross features. However, GDCN introduces GCN, incorporating an information gate at each gated cross layer. It adaptively learns bit-wise gate values to all cross features, enabling fine-grained control over the importance of each cross feature. Notably, both GDCN and DCN-V2 are capable of modeling both bit-wise and vector-wise feature crosses, as demonstrated in DCN-V2.


Although both GDCN and DCN-V2 use the gate mechanism, their purpose and design principles differ. DCN-V2 introduces the idea of MMoE~\cite{ma2018mmoe, shazeer2017outrageously} to decompose the cross matrix into multiple smaller subspaces or "experts." A gating function then combines these experts. This approach primarily reduces non-embedding parameters in its cross matrices while maintaining performance. Differently, GDCN leverages the gating mechanism to choose important cross features adaptively and truly utilize deeper cross features without declining performance. It offers dynamic instance-based interpretability, allowing for a better understanding and analysis of the model's decision-making process. To further enhance the cost-efficiency of GDCN, Section \ref{sec:field dimension optimization} proposes a field-level dimension optimization approach to reduce embedding parameters directly.

\section{Field-level Dimension Optimization}
\label{sec:field dimension optimization}
The embedding dimension typically determines the ability to encode information\cite{zhao2021autodim,sun2021fm2}. However, assigning the same dimension to all fields ignores information capacity in different fields. For example, the number of features in fields like "gender" and "item id" range from $O(2)$ to $O(10^6)$. DCN-V2 and DCN employ a rule-of-thumb formula~\cite{wang2021dcnm, Tensorflow} to assign independent dimensions for each field based on its feature number, i.e., $\left( feature\ number \right) ^{0.25}$. This is a priori method but ignores the true importance of each field. Inspired by FmFM~\cite{sun2021fm2}, we use a posteriori Field-level Dimension Optimization (FDO) method that learns independent dimensions for each field based on its intrinsic importance in a specific dataset.


To begin, we train a full model with a fixed field dimension of 16, as suggested by previous works\cite{sun2021fm2, zhu2022bars, zhu2020fuxictr}. This process allows us to generate an informative embedding table for each field. Next, we employ PCA~\cite{abdi2010principal} to calculate a set of singular values for each field's embedding table, arranged in descending order of magnitude. By evaluating the information utilization (i.e., information ratio), we can determine the optimal dimension by identifying the $argmin_k$ singular values that contribute most significantly to the overall information summation. This step enables us to select a suitable condensed dimension for each field. Lastly, we train a new model with the learned field dimensions from the previous step. In practice, we only need to learn a set of field dimensions once based on a full model and reuse it for subsequent model refreshes.

Table \ref{tab:80_90} presents the optimized dimension for each field with 80\% and 95\% information ratio. When retaining a 95\% ratio, the field dimension ranges from 2 to 15. Decreasing the information ratio results in a reduction in the dimension of each field. Fields with enormous features sometimes require a higher dimension, as observed in fields \{\#23, \#24\}. However, this is not always the case; for instance, fields \{\#16, \#25\} exhibit smaller dimensions. In section ~\ref{exp:dimension optimization}, we present experimental evidence showing that the dimension of a field is strongly correlated with its importance in the prediction process rather than its feature number. Moreover, by preserving over 80\% information ratio, we can obtain lighter GCN models, which slightly surpass the performance of the GCN model with full embedding dimensions and exceed other SOTA models. We also conduct a more comprehensive analysis of FDO to understand the connection between field dimension and its inherent importance.


\begin{table}
    \setlength{\belowcaptionskip}{-0.2cm}
\centering
\caption{Optimized dimension of each field in Criteo dataset (Refer to section \ref{sec_exp}), 95\% and 80\% information ratio.}
\label{tab:80_90}
\scalebox{0.70}{
\begin{tabular}{c|rrr c|rrr} 
\hline
\hline
\multirow{2}{*}{\begin{tabular}[c]{@{}c@{}}Field \\ID\end{tabular}} & \multirow{2}{*}{\begin{tabular}[c]{@{}c@{}}Feature\\Number\end{tabular}} & \multicolumn{2}{c}{\textbf{ Emb Dim}} & \multirow{2}{*}{\begin{tabular}[c]{@{}c@{}}Field \\ID\end{tabular}} & \multirow{2}{*}{\begin{tabular}[c]{@{}c@{}}Feature \\Number\end{tabular}} & \multicolumn{2}{c}{\textbf{\textbf{Emb Dim}}} \\ 
\cline{3-4} \cline{7-8}
 &  & 95\% & 80\% &  &  & 95\% & 80\% \\ 
\hline
\#1 & 49 & 5 & 2        & \#22 & 4 & 2 & 2\\
\#2 & 101 & 13 & 8     & \#23 & 42,646 & 15 & 10 \\
\#3 & 126 & 7 & 3      & \#24 & 5,178 & 14 & 10 \\
\#4 & 45 & 4 & 2       & \#25 & 192,773 & 2 & 2 \\
\#5 & 223 & 13 & 7     & \#26 & 3,175 & 13 & 9 \\
\#6 & 118 & 9 & 4      &  \#27 & 27 & 4 & 3 \\
\#7 & 84 & 8 & 3       &\#28 & 11,422 & 13 & 9 \\
\#8 & 76 & 5 & 2        & \#29 & 181,075 & 5 & 4 \\
\#9 & 95 & 9 & 3       & \#30 & 11 & 5 & 4 \\
\#10 & 9 & 4 & 2        & \#31 & 4,654 & 12 & 8 \\
\#11 & 30 & 3 & 2       &  \#32 & 2,032 & 11 & 7 \\
\#12 & 40 & 4 & 2       &  \#33 & 5 & 3 & 2\\
\#13 & 75 & 4 & 2       & \#34 & 189,657 & 5 & 2\\
\#14 & 1,458 & 12 & 8  & \#35 & 18 & 5 & 4 \\
\#15 & 555 & 12 & 5    & \#36 & 16 & 7 & 4 \\
\#16 & 193,949 & 4 & 3  &  \#37 & 59,697 & 11 & 8\\
\#17 & 138,801 & 8 & 5  & \#38 & 86 & 5 & 3 \\
\#18 & 306 & 10 & 6    & \#39 & 45,571 & 10 & 6 \\
\cline{5-8}
\#19 & 19 & 6 & 4       &\multicolumn{4}{c}{}\\
\#20 & 11,970 & 14 & 10 & \multicolumn{2}{r}{Weighted Ave. $\overline{D}$} & 5.92 & 3.98 \\
\#21 & 634 & 11 & 6    &\multicolumn{2}{r}{Arithmetic Ave. $\overline{K}$} & 7.87 & 4.85 \\
\hline
\hline
\end{tabular}
}
\end{table}

\textbf{Parameter analysis.}
Let $\mathbb{E} = [\mathbf{E}_1,\mathbf{E}_2,...,\mathbf{E}_F]$ represents the embeddings of all features, where $\mathbf{E}_f$ corresponds to the subset of feature representations for the $f^{th}(1\leq f\leq F)$ field. The number of features in the $f^{th}$ field is denoted as $\lvert \mathbf{E}_{f} \rvert$, and the total number of features of a dataset is $T = \sum_{f=1}^{F} {\lvert \mathbf{E}_{f} \rvert}$. Similarly, let $\mathbf{d}=[d_1,d_2,..,d_F]$ represents the different dimensions for each field, where $d_f$ is the dimension of $f^{th}$ field. For an input instance, the arithmetic average dimension is calculated as $\overline{K} = (\sum_{f=1}^{F} d_{f}) / F$, and the output dimension of the embedding layer is determined as $D=F \overline{K}$. Considering all features, the weighted average dimension is denoted as $\overline{D} = (\sum_{f=1}^{F} d_f{\lvert \mathbf{E}_{f} \rvert}) / T $. The total number of embedding parameters is $\mathcal{P}_e = \sum_{f=1}^{F} d_f{\lvert \mathbf{E}_{f} \rvert} = T\overline{D}$. The number of features $T$ is typically massive in a web-scale dataset. For example, in the well-known Criteo dataset, the original feature number is over 30 million with sparsity over 99.99\%~\cite{song2019autoint, wang2021masknet}, with embedding parameters occupying the most portion of the model parameters. Hence, $\overline{D}$ determines the number of embedding parameters, while $\overline{K}$ mainly affects the number of non-embedding parameters, e.g., the cross matrix $\mathbf{W}^{\left( c \right)}\in\mathbb{R}^{F\overline{K} \times F\overline{K}}$ in DCN-V2 and GCN.

By adopting the FDO approach, we can refine the feature dimension by shrinking unnecessary dimensions for some fields to reduce redundant embedding parameters. When using a fixed dimension of 16, the embedding parameters are $16T$. However, after applying FDO with 95\% information ratio, the embedding parameters decrease to $5.92T$, which accounts for only 37\% of the original embedding parameters. If we calculate the field dimension based on the formula (i.e., $d_f =\lvert \mathbf{E}_{f} \rvert^{0.25}$), the weighted average dimension $\overline{D}$ becomes 18.66, resulting in embedding parameters of $18.66T$, which is larger than $16T$. This formula assigns a larger dimension to the field with massive features, overlooking the specific importance of each field. In contrast, FDO is a posteriori method that learns field-level dimensions based on specific information extracted from the trained embedding table. As the field dimension decreases, the arithmetic average dimension $\overline{K}$ also decreases accordingly (e.g., from 16 to 7.87). Thus the non-embedding parameters in the GCN network, i.e., cross matrix $\mathbf{W}^{\left( c \right)}\in\mathbb{R}^{F\overline{K} \times F\overline{K}}$ and gate matrix $\mathbf{W}^{\left( g \right)}\in\mathbb{R}^{F\overline{K} \times F\overline{K}}$ are also reduced naturally. 

\section{EXPERIMENTAL Analysis}
\label{sec:exp_all}

\subsection{Experiment Setup}
\label{sec_exp}

\textbf{Datasets.}
We choose five widely-used datasets to evaluate our proposed methods with other CTR models, i.e., \textbf{Criteo}~\cite{kaggle_criteo, zhu2020fuxictr,wang2022enhancing}
\textbf{Avazu}~\cite{avazu,song2019autoint,meng2021autopi}
\textbf{Malware}~\cite{malware,wang2021masknet}
\textbf{Frappe}~\cite{baltrunas2015frappe, xiao2017attentional, cheng2020adaptive}
\textbf{ML-tag}~\cite{mltag,xiao2017attentional,cheng2020adaptive}
The statistics of these datasets are shown in Table \ref{Tab.dataset} and detailed descriptions can be found in the given references.

\textbf{Data preparation.} Firstly, we randomly split each dataset into training(80\%), validation (10\%) and testing (10\%) datasets. Secondly, in Criteo and Avazu, we remove infrequent features in a certain field appearing less than \textit{threshold} times and treat them as a dummy feature "\textit{<unkonwn>}". The \textit{threshold} is set to \{10, 5\} for Criteo and Avazu, respectively. Finally, in the Criteo dataset, we normalize numerical values by transforming a value $z$ to $\lfloor log^2(z) \rfloor$ when $x>2$, which is adopted by the winner of the Criteo competition~\cite{winner}.

\begin{table}[t]
    \setlength{\belowcaptionskip}{-0.2cm}
 \centering
 \caption{Dataset statistics. K means thousand.}
 \scalebox{0.80}{
 \begin{tabular}{c|cccc|cc}
 \hline
 \hline
Datasets &Positive& \#Training & \#Validation & \#Testing  & \#Features &\#Fields \\ \hline
Criteo &26\% & 36,672K& 4,584K & 4,584K &1,086K &39\\ 
Avazu &17\% & 32,343K& 4,043K&  4,043K  & 1,544K &23\\ 
Malware &50\%  &7,137K &892K  &892K&976K &81\\
Frappe &33\%  &231K & 29K & 29K & 5K &10\\ 
ML-tag  &33\%  &1,605K  &201K  &201K &90K &3\\ 
 \hline
 \hline
 \end{tabular}
 }
 \label{Tab.dataset}
\end{table}

\textbf{Evaluation Metrics.}
We adopt \textbf{AUC} (Area Under ROC) and \textbf{Logloss} (Cross Entropy) to assess the performance of all models on the testing set. A higher AUC or a lower Logloss (LL) at \textit{0.001-level} can be considered a significant improvement for the CTR prediction task, which is the common consensus in existing works~\cite{wang2022enhancing, zhu2022nasctr, wang2021dcnm, huang2019fibinet, lian2018xdeepfm}. Additionally, $\Delta{AUC}$ and $\Delta{LL}$ are calculated to show the averaged performance improvement compared to a given benchmark over five datasets in Table \ref{tab:all} and \ref{tab:parallel}. Also, $\textit{Rel.Imp}$ denotes the relative improvements compared with the best baseline. 

\textbf{Comparison methods.}
To evaluate our methods, we compare them with four classes of representative methods. 
1) First-order method, e.g., \textbf{LR}~\cite{richardson2007predicting};
2) FM-based methods that model second-order cross features, including \textbf{FM}~\cite{rendle2012factorization}, 
\textbf{FwFM}~\cite{pan2018field}, 
\textbf{DIFM}~\cite{lu2020dual}, and
\textbf{FmFM}~\cite{sun2021fm2};
3) Methods that capture high-order cross features, including 
\textbf{CrossNet(CN)}~\cite{wang2017deep}, 
\textbf{CIN}~\cite{lian2018xdeepfm},
\textbf{AutoInt}~\cite{song2019autoint}, 
\textbf{AFN}~\cite{cheng2020adaptive}, 
\textbf{CN-V2}~\cite{wang2021dcnm},
\textbf{IPNN}~\cite{qu2018product}, 
\textbf{OPNN}~\cite{qu2018product},
\textbf{FINT}~\cite{zhao2021fint}, 
\textbf{FiBiNET}~\cite{huang2019fibinet} and 
\textbf{SerMaskNet}~\cite{wang2021masknet}.  
4) representative ensemble/parallel methods, including 
\textbf{WDL}~\cite{cheng2016wide}, 
\textbf{DeepFM}~\cite{guo2017deepfm}, 
\textbf{DCN}~\cite{wang2017deep}, 
\textbf{xDeepFM}~\cite{lian2018xdeepfm}, 
\textbf{AutoInt+}~\cite{song2019autoint}, 
\textbf{AFN+}~\cite{cheng2020adaptive}, 
\textbf{DCN-V2}~\cite{wang2021dcnm},
\textbf{NON}~\cite{luo2020network},
\textbf{FED}~\cite{zhao2020fed}, 
and \textbf{ParaMaskNet}~\cite{wang2021masknet}. 
We do not show the results of some methods, e.g., CCPM~\cite{gehring2017convolutional}, GBDT~\cite{chen2016xgboost}, FFM~\cite{juan2016field}, HoFM~\cite{blondel2016higher}, AFM~\cite{xiao2017attentional}, NFM~\cite{he2017neural}, as many works~\cite{cheng2020adaptive, wang2021dcnm, wang2021masknet} have surpassed them.

\begin{table*}
    \setlength{\abovecaptionskip}{0.2cm}
    \setlength{\belowcaptionskip}{-0.2cm}
\centering
\caption{Overall performance comparison in the five datasets. $\Delta{AUC}$ and $\Delta{LL}$ indicate averaged performance boost compared with SerMaskNet. Bold scores are the best performance, and underlined scores are the best baseline performance. }
\label{tab:all}
\scalebox{0.85}{
\begin{tabular}{cc|cc|cc|cc|cc|cc|cc} 
\hline 
\hline

\multirow{2}{*}{\begin{tabular}[c]{@{}c@{}}Model\\Class\end{tabular}}
& Datasets& \multicolumn{2}{c|}{Criteo} & \multicolumn{2}{c|}{Avazu} & \multicolumn{2}{c|}{Malware} &\multicolumn{2}{c|}{Frappe}& \multicolumn{2}{c|}{ML-tag}&
\multirow{2}{*}{\begin{tabular}[c]{@{}c@{}}$\Delta{AUC} $\\ $\uparrow$ \end{tabular}} &
\multirow{2}{*}{\begin{tabular}[c]{@{}c@{}}$\Delta{LL} $\\$\downarrow$\end{tabular}} \\

\cline{2-12}
& Model& AUC& Logloss& AUC& Logloss& AUC& Logloss& AUC&Logloss& AUC& \multicolumn{1}{c|}{Logloss} & \multicolumn{1}{l}{}& \multicolumn{1}{l}{}\\ 
\hline
\multirow{1}{*}{\begin{tabular}[c]{@{}c@{}}First-order\end{tabular}} 
& LR  &0.7937 &0.4562 &0.7573 & 0.3925  &0.7107 & 0.6196  &0.9376&0.2882&0.9339 &0.2956   &-3.62\% & 23.87\%\\
\hline
\multirow{4}{*}{\begin{tabular}[c]{@{}c@{}}Second-\\order\end{tabular}} 
& FM    &0.8085&0.4433&0.7829&0.3785 &0.7363 &0.5982&0.9583&0.2336&0.9539 &0.2523 &-1.08\%	& 11.33\%\\
& FwFM  &0.8112&0.4408&0.7857&0.3772 &0.7367 &0.5980&0.9738&0.1851&0.9591 &0.2307 &-0.51\%	&3.30\% \\
& DIFM  &0.8128&0.4395&0.7887&0.3748 &0.7424 &0.5926&0.9792&0.1880&0.9569 &0.2316 &-0.18\%	&3.37\%\\
& FmFM  &0.8122&0.4399&0.7874&0.3765 &0.7433 &0.5922&0.9762&0.1875&0.9589 &0.2357 &-0.22\%	&3.76\%\\ 
\hline

\multirow{10}{*}{\begin{tabular}[c]{@{}c@{}}High-\\order\end{tabular}} 
&CN      &0.8071&0.4442&0.7853 &0.3784	&0.7265	&0.6049	&0.9772	&0.1876	&0.9498	&0.2582 &-1.02\%	&6.45\% \\
&CIN     &0.8121&0.4398&0.7881 &0.3754  &0.7429 &0.5924 &0.9792 &0.1850 &0.9593 &0.2490 &-0.15\%	&4.55\%   \\
&AutoInt &0.8118&0.4399&0.7881 &0.3750  &0.7363 &0.5976 &0.9788 &0.1671 &0.9540 &0.2575 &-0.45\%	&3.25\%  \\
&AFN     &0.8116&0.4401&0.7883 &0.3752  &0.7427 &0.5924 &0.9801 &0.1674 &0.9587 &0.2305 &-0.15\%	&0.78\%  \\
&CN-V2   &0.8140&0.4380&\underline{0.7893} &\underline{0.3745}&0.7383&0.5959 &0.9803&0.1710&0.9549&0.2529 &-0.26\%	&3.16\% \\
&IPNN    &0.8128&0.4390&0.7890&0.3747 &0.7433 & 0.5918  &0.9801&0.1724 &0.9598 &0.2344  &-0.07\%	&1.64\%  \\
&OPNN    &0.8135&0.4383&0.7892 &0.3746&0.7436 & 0.5918  &0.9804&0.1645 &0.9595 &0.2346  &-0.04\%	&0.65\% \\
&FiBiNet &0.8129&0.4389&0.7889 &0.3746&\underline{0.7441}&\underline{0.5914}&0.9789 &0.1777&0.9595&0.2352&-0.08\%	&2.33\%\\
&FINT    &0.8128& 0.4390 &0.7891 &0.3747  &0.7424&0.5924& \underline{0.9807}&0.1631 &0.9598 & 0.2356 &-0.08\%	&0.62\%\\
&SerMaskNet &\underline{0.8141}&\underline{0.4379}&0.7891&0.3746&0.7440&0.5920&0.9804 &\underline{0.1628}&\underline{0.9602} &\underline{0.2297} & - 	&-\\
\hline
\multirow{3}{*}{\begin{tabular}[c]{@{}c@{}}Ours\end{tabular}} 
& GCN  &$0.8154^{\star}$ & $0.4367^{\star}$&$0.7903^{\star}$&$ 0.3742^{\star}$&$0.7445^{\star}$&$0.5908^{\star}$&$0.9820^{\star}$&$0.1634^{\star}$&$0.9619^{\star}$&$0.2250^{\star}$&0.14\%	&-0.45\% \\
& GDCN-S &$\textbf{0.8158}^{\star}$ &$\textbf{0.4364}^{\star}$ &$\textbf{0.7905}^{\star}$&$\textbf{0.3739}^{\star}$ &$\textbf{0.7456}^{\star}$ & $\textbf{0.5899}^{\star}$ &$\textbf{0.9838}^{\star}$&$\textbf{0.1470}^{\star}$&$\textbf{0.9645}^{\star}$&$\textbf{0.2230}^{\star}$&\textbf{0.28\%}	&\textbf{-2.70\%} \\
  & \textit{Rel.Imp}   &0.0017&-0.0015&0.0012&-0.0006 &0.0015&-0.0015   &0.0031&-0.0158 &0.0043&-0.0067 &-&- \\
\hline
\bottomrule
\end{tabular}
}
\end{table*}

\textbf{Implementation details.}
We implement all models with Pytorch~\cite{paszke2019pytorch} and refer to existing works~\cite{zhu2022bars, zhu2020fuxictr, chen2021dcap}. We use Adam~\cite{kingma2014adam} optimizer to optimize all models, and the default learning rate is 0.001. We utilize the Reduce-LR-On-Plateau scheduler during the training process to reduce the learning rate by 10 when the performance stops improving in 3 consecutive epochs. We apply early stopping with the patience of 5 on the validation set to avoid overfitting. The batch size is set to 4096. The embedding dimension for all datasets is 16. Following previous works~\cite{guo2017deepfm, cheng2020adaptive, meng2021autopi, song2019autoint, huang2019fibinet, chen2021enhancing}, we adopt the same structures (i.e., 3 layers, 400-400-400) for models that involve DNN for a fair comparison. Unless otherwise specified, all activation functions are ReLU, and the dropout rate is 0.5. For our proposed GCN, GDCN-S and GDCN-P, the default number of gated cross layer is 3 without specifically mentioned. For other baselines, we refer to two benchmark works (i.e., BARS~\cite{zhu2022bars} and FuxiCTR~\cite{zhu2020fuxictr}) and their original literature to finetune their hyper-parameters.

\textbf{Significance Test.} To ensure a fair comparison, we run each method \textbf{10 times} with random seeds on a single GPU (NVIDIA TITAN V) and report the average testing performance. We perform a two-tailed t-test~\cite{liu2019fgcnn,wang2022mcrf,wang2022enhancing} to detect the statistical significance between our method and the best baseline methods. The improvements over the best baselines are \textbf{statistically significant with p<0.01} in all experiments, represented by $\star$ in Table \ref{tab:all} and Table \ref{tab:parallel}. 



\subsection{Overall Performance}
\subsubsection{Comparison within stacked models} 
We compare GCN and GDCN-S with stacked baseline models, including first-order, second-order and high-order models. The overall performance is summarized in Table \ref{tab:all}. We have the following observations:
    
First, in most cases, high-order models outperform first- and second-order models, demonstrating the effectiveness of learning complex high-order feature interactions. Notably, models such as OPNN, FiBiNet, FINT and SerMaskNet perform even better, which simultaneously capture explicit and implicit feature crosses with a stacked DNN. This confirms the rationale behind modeling both explicit and implicit high-order feature interactions. 

Second, GCN consistently outperforms all stacked baseline models by considering only explicit polynomial feature interactions. GCN is a generalization of CN-V2, with the addition of an information gate to identify meaningful cross features. The performance of GCN validates that not all cross features are beneficial to the final prediction, and a large number of irrelevant interactions introduce unnecessary noise. By adaptively re-weighting cross features in each order, GCN achieves significant performance improvement over CN-V2. Moreover, it outperforms SerMaskNet by increasing the average $\Delta{AUC}$ by 0.14\% and improving $\Delta{LL}$ by 0.45\%.

Third, GDCN-S surpasses all stacked baselines and achieves the best performance. In GDCN-S, the stacked DNN further learn implicit interaction information over the GCN structure. As a result,  GDCN-S outperforms GCN and achieves superior prediction accuracy compared to other stacked models, e.g., OPNN, FINT and SerMaskNet. Specifically, GDCN-S achieves an average improvement of 0.28\% ($\Delta{AUC}$) and 2.70\% ($\Delta{LL}$) compared to SerMaskNet. 

\begin{table*}
    \setlength{\abovecaptionskip}{0.2cm}
    \setlength{\belowcaptionskip}{-0.2cm}
\centering
\caption{Performance of parallel models, which integrates implicit feature interactions, i.e., DNN. We list the main parallel networks for each model. "Att." and "Concat." represents the attention and concatenation operation, respectively.}
\label{tab:parallel}
\scalebox{0.80}{
\begin{tabular}{cc|cc|cc|cc|cc|cc|cc} 
\hline
\hline
\multicolumn{2}{c|}{Datasets}  & \multicolumn{2}{c|}{Criteo} & \multicolumn{2}{c|}{Avazu}& \multicolumn{2}{c|}{Malware} & \multicolumn{2}{c|}{Frappe} & \multicolumn{2}{c|}{ML-tag} & \multirow{2}{*}{\begin{tabular}[c]{@{}c@{}}$\Delta{AUC} $\\ $\uparrow$\end{tabular}} & \multicolumn{1}{c}{\multirow{2}{*}{\begin{tabular}[c]{@{}c@{}}$\Delta{LL} $\\$\downarrow$\end{tabular}}} \\ 
\cline{1-12}
Model & Networks & AUC & Logloss & AUC & Logloss& AUC & Logloss & AUC & Logloss & AUC & Logloss &  & \\ 
\hline
DNN & DNN & 0.8139 & 0.4383 & 0.7886 & 0.3748 &0.7424 & 0.5928& 0.9794 & 0.1661 & 0.9598 & 0.2418 & - &-  \\
\hline
WDL & LR, DNN & 0.8141 & 0.4379 & 0.7888 & 0.3747 &0.7436 & 0.5915& 0.9801 & 0.1652 & 0.9605 & 0.2411 & 0.07\%	&-0.23\%  \\
DeepFM&FM, DNN & 0.8140 & 0.4378 & 0.7891 & 0.3745 &0.7424 & 0.5925& 0.9808 & 0.1613 & 0.9606 & 0.2440 &  0.05\%	&-0.42\%  \\
DCN&CN, DNN & 0.8142 & 0.4378 & 0.7890 & 0.3745 &0.7434&0.5917& 0.9805 & 0.1618 & 0.9607 & 0.2283 &  0.08\%	&-1.69\%  \\
xDeepFM &CIN, DNN & 0.8142 & 0.4377 & 0.7893 & 0.3744 &\underline{0.7442} & \underline{0.5911} & 0.9806 & 0.1638 & 0.9609 & 0.2328 &  0.12\%	&-1.10\%  \\
AutoInt+& AutoInt, DNN & \underline{0.8144} & 0.4376 & 0.7886 & 0.3745 &0.7435 & 0.5917& 0.9802 & 0.1654 & 0.9609 & 0.2343 &  0.07\%	&-0.76\%  \\
AFN+& AFN, DNN & 0.8138 & 0.4384 & 0.7887 & 0.3748 &0.7439&0.5912& 0.9807 & 0.1620 & 0.9605 & 0.2295 &  0.07\%	&-1.53\%  \\
DCN-V2&CN-V2, DNN &\underline{0.8144} & \underline{0.4375} & \underline{0.7898} & \underline{0.3743} &0.7433 & 0.5920&0.9810 & \underline{0.1539} & 0.9603 & 0.2284 &  0.11\%	&-2.67\%  \\ 

\hline
NON&LR, Att., DNN  & 0.8133 & 0.4390 & 0.7889 & 0.3747&0.7426 & 0.5925 & 0.9803 & 0.1564 &0.9609	&0.2279 &  0.03\%	&-2.26\%  \\
FED&Concat., Att.,  DNN & 0.8141 & 0.4379 & 0.7891 & 0.3745&0.7436 & 0.5915&\underline{0.9811} & 0.1590 & \underline{0.9610} & 0.2294 & 0.12\%	&-1.99\%  \\ 
ParaMaskNet&MaskBlocks, DNN &0.8142&0.4377  &0.7895 & 0.3748  &0.7435& 0.5920  &0.9805 &0.1559   &0.9608 &\underline{0.2272}   &0.10\% 	&-2.47\%\\
\hline
GDCN-P&GCN, DNN &$\textbf{0.8161}^{\star}$ & $\textbf{0.4360}^{\star}$ & $\textbf{0.7909}^{\star}$ &$ \textbf{0.3733}^{\star}$&$\textbf{0.7462}^{\star}$ &$\textbf{0.5893}^{\star}$&$\textbf{0.9852}^{\star}$ & $\textbf{0.1410}^{\star}$ & $\textbf{0.9663}^{\star}$ & $\textbf{0.2144}^{\star}$ &  \textbf{0.46\%}	&\textbf{-5.57\%}  \\
\multicolumn{2}{c|}{\textit{Rel.Imp}}  & 0.0017 & -0.0015 & 0.0011 & -0.0010 &0.0020&-0.0018& 0.0041 & -0.0129 & 0.0053 & -0.0128 &- & -\\

\hline
\bottomrule
\end{tabular}
}
\end{table*}

\subsubsection{Comparison within parallel models}
Table \ref{tab:parallel} presents the performance of SOTA ensemble/parallel models. Each method incorporates parallel networks, such as FM and DNN in DeepFM, and CN-V2 and DNN in DCN-V2. Additionally, we compare these models with the vanilla DNN model and calculate the average improvement $\Delta{AUC}$ and $\Delta{LL}$ based on it. Our observations are as follows:

Firstly, integrating implicit and explicit feature interactions enhances predictive ability. While DNN solely models implicit feature interactions, parallel models that combine DNN with other networks to capture explicit feature interactions outperform individual networks. Notably, GCN shows the most substantial average performance improvement, as evident from the $\Delta{AUC}$ and $\Delta{LL}$ values of 0.46\% and 5.57\%, respectively. This highlights the superior power of GCN compared to other basic operation networks.

Secondly, GDCN-P outperforms all parallel and stacked models. When compared to the parallel models in Table \ref{tab:parallel}, GDCN-P achieves superior performance, surpassing all the best baselines indicated with \textit{Rel.Imp} of 0.0011 to 0.0053 on AUC and -0.0010 to -0.0129 on Logloss. Furthermore, when considering stacked models and using SerMaskNet from Table \ref{tab:all} as the benchmark, GDCN-P achieves an average improvement of 0.38\% ($\Delta{AUC}$) and 4.26\% ($\Delta{LL}$).


\subsection{Deeper High-order Feature Crossing}
\label{sec:high_order}

\subsubsection{Compare GCN to other models by changing cross depth.} 
\begin{figure}[t]
\setlength{\abovecaptionskip}{0.2cm}
\setlength{\belowcaptionskip}{-0.2cm}
\centering
\includegraphics[width=0.38\textwidth]{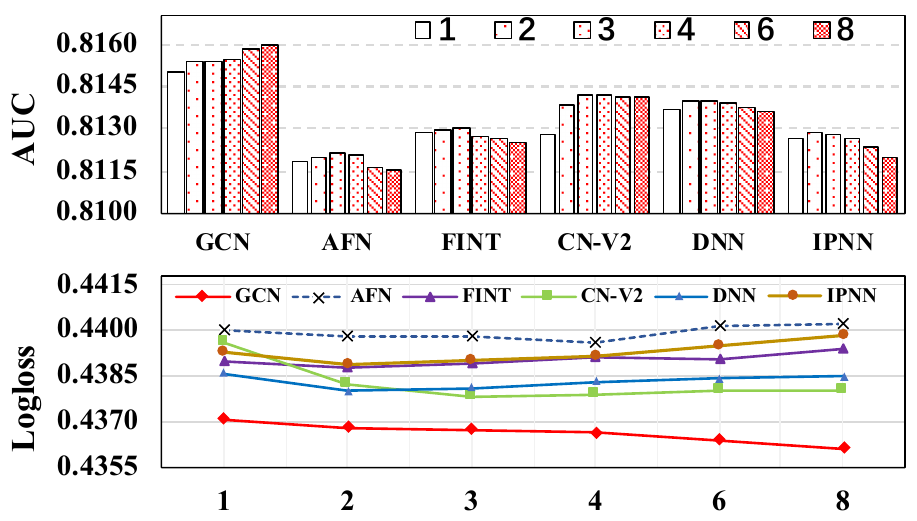}
\caption{Comparison of different cross depth on six models.}
\label{fig:depth}
\end{figure}

We compare GCN with five representative models that can capture high-order interactions, namely AFN~\cite{cheng2020adaptive}, FINT~\cite{zhao2021fint}, CN-V2~\cite{wang2021dcnm}, DNN~\cite{zhang2016fnn} and IPNN~\cite{qu2018product}. Figure \ref{fig:depth} illustrates the testing AUC and Logloss as the cross depth increases on the Criteo dataset. 

As the cross layers increase, the performance of the five compared models improves. However, their performance experiences a significant decline when the cross depth goes deeper (e.g., over 2, 3, or 4 cross layers). These models can capture deeper high-order explicit and implicit feature interactions functionally, but higher-order cross features also introduce unnecessary noise, which can result in overfitting and lead to a degradation in results. This issue is also observed in the hyper-parameter analysis of cross depth conducted in many SOTA works \cite{lian2018xdeepfm, cheng2020adaptive, blondel2016higher, song2019autoint, wang2017deep, wang2021dcnm, guo2017deepfm, qu2018product, zhao2021fint, covington2016youtobednn}.

In contrast, the performance of GCN continues to improve as the number of cross layers increases from 1 to 8. Specifically, GCN incorporates an information gate compared to the CN-V2 model. This information gate enables GCN to identify useful cross features in each order and accumulate only necessary information for final prediction. Therefore, GCN consistently outperforms CN-V2 and other models, even when the cross depth becomes deeper, verifying the design rationality of the information gate.

\subsubsection{The performance of GCN and GDCN-P with deeper gated cross layers.}
\begin{figure}[t]
\setlength{\abovecaptionskip}{0.2cm}
\setlength{\belowcaptionskip}{-0.2cm}
\centering
\includegraphics[width=0.42\textwidth]{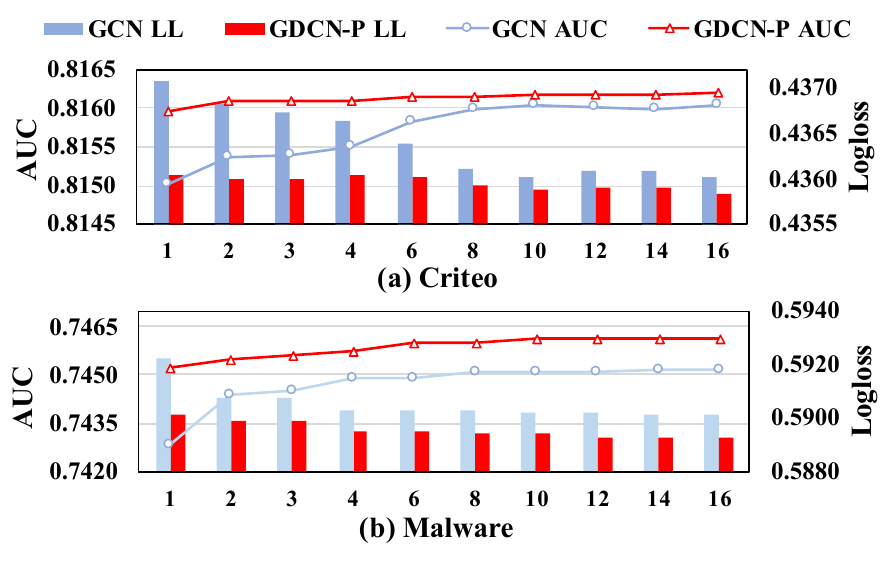}
\caption{Performance with deeper gated cross layers.} 
\label{fig:deeper_layers}
\end{figure}

We further increase the cross depth of GCN and GDCN-P on the Criteo and Malware datasets from 1 to 16. Notably, we kept the depth of the DNN fixed at 3 in GDCN-P to prevent overfitting caused by the DNN. Figure \ref{fig:deeper_layers} presents the experimental results.

As the cross depth increases, the performance of both GCN and GDCN-P improves, and the trend observed in GDCN-P is consistent with that of GCN. GDCN-P consistently outperforms GCN after incorporating the DNN component, highlighting the importance of capturing implicit feature interactions using DNN. Furthermore, DNN enables GDCN-P to achieve the best results earlier. Specifically, the performance of GCN reaches a plateau when the depth exceeds 8 or 10, whereas the threshold for GDCN-P is 4 or 6. Although GCN and GDCN-P can select valuable cross features, the usefulness of high-order cross features decreases sharply as the cross depth increases. This phenomenon aligns with the common intuition that individuals are not typically influenced by too many features simultaneously when making a decision, such as clicking or purchasing an item. Notably, our models' performance remains stable instead of decreasing after surpassing the plateau threshold.

\subsection{The Interpretability of GCN}

Interpretability is vital in understanding the reasoning behind specific predictions and enhancing confidence in the prediction results. The GCN offers both static and dynamic interpretations from the model and instance perspectives, facilitating a comprehensive understanding of the model's decision-making process.

\subsubsection{Static model interpretability.}
Within GCN, the cross matrix $\mathbf{W}^{(c)}$ serves as an indicator of the relative importance of interactions among different fields. If each instance consists of $F$ fields with the same field size $d$, the cross matrix can be represented in both a bit-wise and block-wise manner, as demonstrated in Equ.\ref{equ:bit_block}.
\begin{align}
\textstyle
\label{equ:bit_block}
\mathbf{W}^{\left( c \right)}=\left[ \begin{matrix}
	w_{1,1}&		\cdots&		w_{1,Fd}\\
	\vdots&		\ddots&		\vdots\\
	w_{Fd,1}&		\cdots&		w_{Fd,Fd}\\
\end{matrix} \right] =\left[ \begin{matrix}
	W_{1,1}&		\cdots&		W_{1,F}\\
	\vdots&		\ddots&		\vdots\\
	W_{F,1}&		\cdots&		W_{F,F}\\
\end{matrix} \right] 
\end{align}
Each block-wise matrix $W_{i,j} \in \mathbb{R}^{d\times d}$ shows the importance of the 1st-order cross between the i-th and j-th fields. When FDO is applied to learn various field dimensions, $d_i$ and $d_j$ in block $W_{i,j} \in \mathbb{R}^{d_i\times d_j}$ are the different dimensions of the i-th and j-th fields. Additionally, as cross layers increase, the corresponding cross matrix can further indicate the inherent relation importance among multiple fields. 

\begin{figure}[tb]
\setlength{\abovecaptionskip}{0.2cm}
\setlength{\belowcaptionskip}{-0.2cm}
\centering
\subfloat[GCN]{
\begin{minipage}[t]{0.52\linewidth}
\centering
\includegraphics[width=1.0\textwidth]{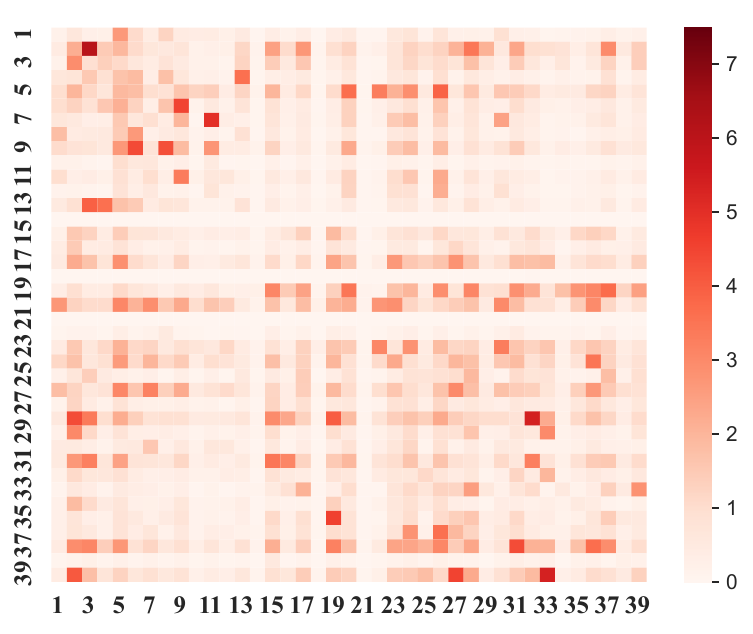}
\end{minipage}
}
\subfloat[GCN-FDO (95\%)]{
\begin{minipage}[t]{0.52\linewidth}
\centering
\includegraphics[width=1.0\textwidth]{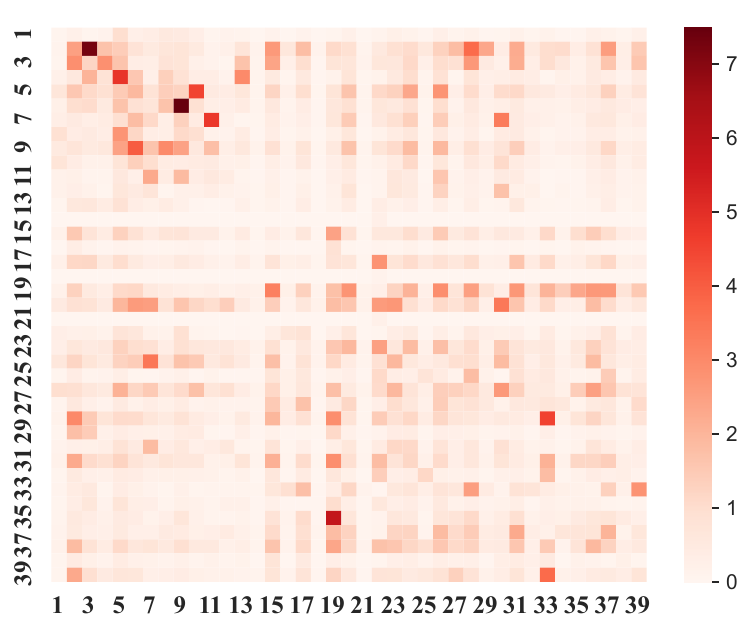}
\end{minipage}
}

\centering
\caption{Visualization of the block-wise cross matrix $\mathbf{W}_1^{(c)}$ in 1st gated cross layer learned by GCN and GCN-FDO (95\%). } 
\label{fig:heatmap_cross}
\end{figure}

Figure~\ref{fig:heatmap_cross}(a) shows the heatmap of the block-wise cross matrix $\mathbf{W}_1^{(c)}$ in 1st cross layer learned by GCN on the Criteo dataset. Similar to DCN-V2~\cite{wang2021dcnm}, each color box represents the \textit{Frobenius norm} of the corresponding block-wise matrix $W_{i,j}$, indicating the importance of field crosses. Darker shades of red indicate stronger learned crosses, such as <\#3, \#2> and <\#9, \#6>. When applying the FDO approach to GCN, GCN-FDO still captures similar cross importance, as depicted in Figure \ref{fig:heatmap_cross}(b). The cosine similarity between the two matrices is 0.89, demonstrating a strong consistency in capturing the inherent field cross importance before and after the application of FDO. 

\subsubsection{Dynamic instance interpretability.}

Model-based interpretability, which captures static relationships among different fields, has limitations as the cross matrices remain fixed once a model is trained. However, GCN also offers dynamic interpretability through gate weights learned by the information gate, providing both bit-wise and field-wise interpretations for each input instance.

\begin{figure}[tb]
\setlength{\abovecaptionskip}{0.2cm}
\setlength{\belowcaptionskip}{-0.2cm}
\centering
\subfloat[Bit-wise gate value vectors.]{
\begin{minipage}[t]{0.99\linewidth}
\centering
\includegraphics[width=1.0\textwidth]{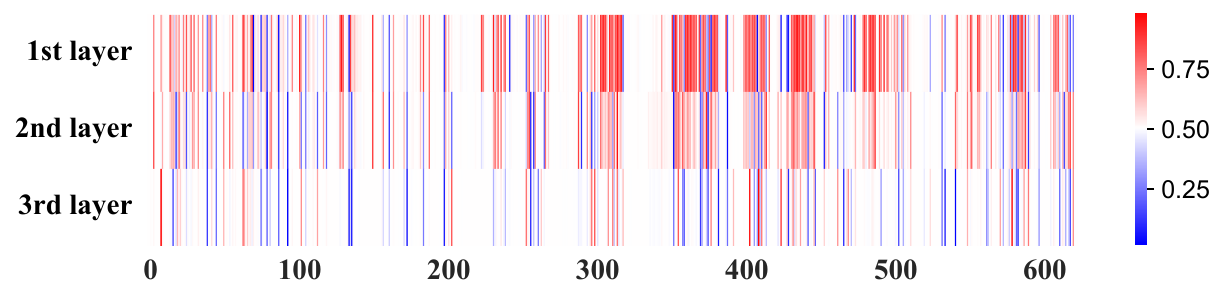}
\end{minipage}
}

\subfloat[Field-wise gate value vectors.]{
\begin{minipage}[t]{0.99\linewidth}
\centering
\includegraphics[width=1.0\textwidth]{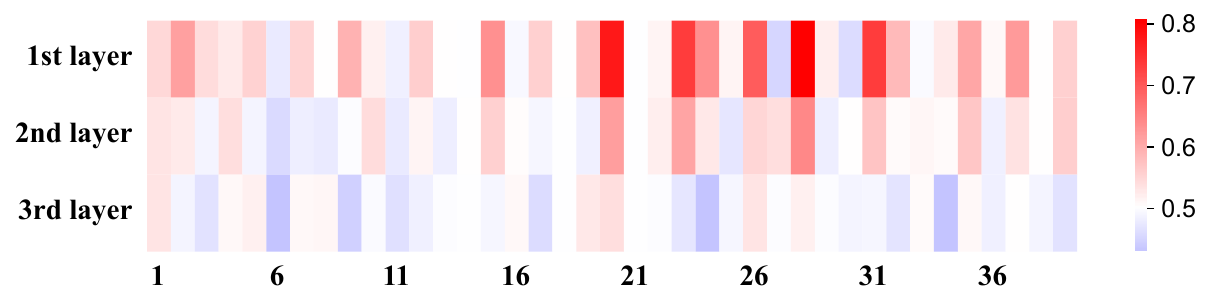}
\end{minipage}
}
\centering
\caption{Visualization of the gate value vectors. }
\label{fig:gate_value}
\end{figure}

We randomly select two instances from the Criteo dataset to visualize the learned gate weights and examine the gate values from the 1st to 3rd gated cross layer. Figure \ref{fig:gate_value}(a) presents the bit-wise gate weight vectors, with dimensions of $\mathbb{R}^{1\times(39*16)}$ in each layer, showing the importance of each bit-wise cross. Using the bit-wise gate vector, we derive the field-wise gate vectors by averaging the 16-bit values corresponding to each field. Figure~\ref{fig:gate_value}(b) displays the field-wise gate weight vectors ($\mathbb{R}^{1\times39}$), indicating the importance of each specific feature. As the gate weights are calculated using the Sigmoid function in the gated cross layer, the red blocks (greater than 0.5) indicate important features, while the blue blocks (less than 0.5) indicate unimportant features.

Figures~\ref{fig:gate_value}(a) and \ref{fig:gate_value}(b) reveal the importance of cross features at both the bit-level and field-level and how they vary across cross layers for each instance. Generally, lower-order feature crosses contain more significant features, while higher-order feature crosses contain less important features. In the 1st layer, numerous feature crosses are identified as important (red blocks), whereas in the 2nd and 3rd cross layers, most crosses are neutral (white blocks) or unimportant (blue blocks), particularly in the 3rd layer. This aligns with our intuition: as the cross layer increases, the number of important cross features significantly decreases, so we design the information gate to choose the important features adaptively. In contrast, most models fail to select useful feature crosses when modeling high-order crosses, resulting in a performance drop, as confirmed in Section~\ref{sec:high_order}. Moreover, from Figure~\ref{fig:gate_value}(b), we can identify specific features that are important or unimportant, such as features \{\#20, \#23, \#28\} being important and features \{\#6, \#11, \#30\} being unimportant. We can also refer to the names of these specific important features from the input instance. Once we know which features are influential, we can interpret and even intervene in relevant features that contribute to the click probability for users.

\begin{figure}[t]
\setlength{\abovecaptionskip}{0.2cm}
\setlength{\belowcaptionskip}{-0.2cm}
\centering
\includegraphics[width=0.50\textwidth]{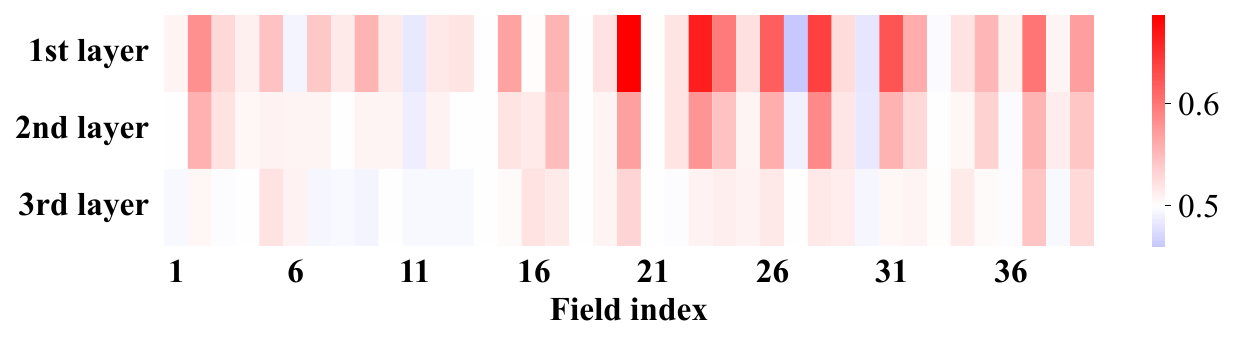}
\caption{Visualization of average field-wise gate vectors on 1,000,000 instances.}
\label{fig:100w_gate values}
\end{figure}

Lastly, in Figure~\ref{fig:100w_gate values}, we record and average the field-level gate vectors from 1,000,000 instances, indicating the average importance of each field, particularly in the 1st layer. For example, fields \{\#20, \#23, \#24\} are significantly important, while fields \{\#11, \#27, \#30\} are relatively less important. Furthermore, Figure~\ref{fig:100w_gate values} further verifies the significant decrease in the number of important cross features as the cross layer increases from a statistical perspective.

\subsection{Comprehensive Analysis of FDO}
\label{exp:dimension optimization}

\subsubsection{Effectiveness analysis.} We apply the FDO approach to GCN and assess its performance by keeping various information ratio from 50\% to 98\%. Table \ref{tab:variance} shows the results, including the weight average dimension $\overline{D}$, arithmetic dimension $\overline{K}$ and the total number of parameters (\#Params). When maintaining 80\% ratio, we only require 23\% of the parameters, while the model's performance is on par with the full model. Moreover, when the information ratio is between 80\% to 98\%, the model performance is even better than the full model. This is because FDO refines the feature dimensions for each field, eliminating non-essential dimensional information and reducing redundancy during the cross process. However, when the information ratio is lower (i.e., less than 80\%), the shorter dimension fails to adequately represent the features in the corresponding field, resulting in a significant decline in prediction performance. Lastly, we reduce the full model's dimension from 16 to 8, approximately equivalent to the weighted average dimension (7.56) with a 95\% ratio. Although it directly decreases the model's parameters, it also leads to a drop in performance. Other studies~\cite{song2019autoint, he2017ncf, he2017neural, sun2020generic} have likewise confirmed the negative impact of reducing embedding size. In comparison, the FDO method can help reduce parameters number more directly while achieving comparable performance.

\begin{table}[t]
\setlength{\abovecaptionskip}{0.2cm}
\setlength{\belowcaptionskip}{-0.2cm}
\centering
\caption{The effectiveness of GCN with FDO approach. All results are statistically significant with p<0.001 compared to the full (16) model with fixed dimension.  M is million.} 
\label{tab:variance}
\scalebox{0.88}{
\begin{tabular}{c|cc|cc|cc}
\hline
\hline
\multirow{1}{*}{\begin{tabular}[c]{@{}c@{}}Ratio\end{tabular}} & \multirow{1}{*}{$\overline{D}$} & \multirow{1}{*}{$\overline{K}$} & \multicolumn{1}{c}{\multirow{1}{*}{\begin{tabular}[c]{@{}c@{}}\#Params\end{tabular}}}&\multicolumn{1}{c|}{Proportion} & \multicolumn{1}{c}{AUC} &\multicolumn{1}{c}{Logloss}  \\ 
\hline
Full (16) & 16  & 16  & 19.73M&100\%&0.8154 &0.4367\\
98\%& 7.56& 9.54& 9.04M&45.8\%&0.8157 &0.4365\\
95\%& 5.92& 7.87& 7.00M&35.5\%&0.8157 &0.4365\\
90\%& 4.99& 5.92& 5.80M&29.4\%&0.8156 &0.4365\\
80\%& 3.98& 4.85& 4.54M&23.0\%&0.8155 &0.4366\\
70\%& 2.94& 3.69& 3.32M&16.8\%&0.8152 &0.4371\\
60\%& 2.41& 2.90& 2.69M&13.6\%&0.8146 &0.4374\\ 
50\%& 1.97& 2.26& 2.54M&11.1\%&0.8137 &0.4380\\
\hline
\hline
Full (8)& 8  & 8  & 9.27M&47.9\%&0.8151 &0.4371\\
\hline\hline
\end{tabular}}
\end{table}

\subsubsection{Memory and efficiency comparison.} 
\begin{figure}[t]
\setlength{\abovecaptionskip}{0.2cm}
\setlength{\belowcaptionskip}{-0.2cm}
    \centering
    \includegraphics[width=0.41\textwidth]{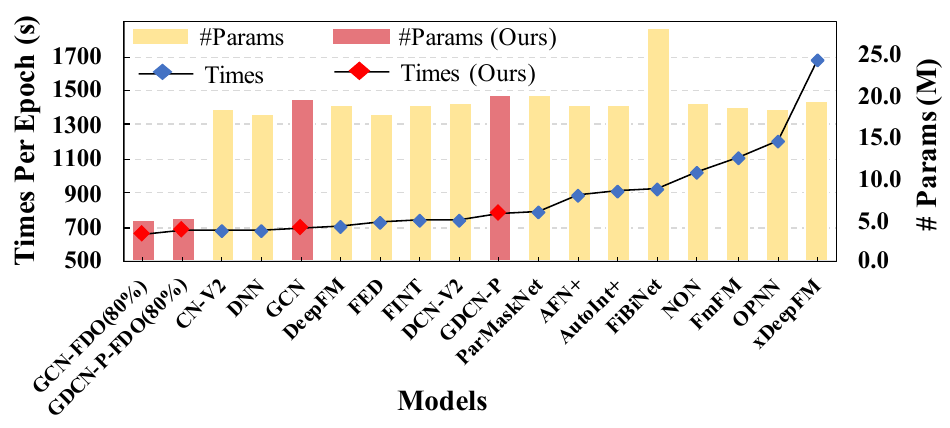}
    \caption{The comparison of model parameters (\#Params) and the training time per epoch on Criteo dataset.}
    \label{fig:auc_params}
\end{figure}

On the Criteo dataset, we have a total of 1.086M features after removing infrequent ones. When using a fixed dimension (K=16), the embedding parameters are around 17.4M. As shown in Figure~\ref{fig:auc_params}, most existing models have parameters ranging from 18M to 20M, primarily due to the embedding parameters. By applying FDO, we can directly reduce the embedding parameters. Meanwhile, GCN-FDO and GDCN-P-FDO achieve comparable accuracy with only around 5M model parameters. We further compare the training time among existing SOTA models. The training time of the GCN and GDCN without dimension optimization is comparable to that of DNN and DCN-V2. After applying FDO, GCN-FDO and GDCN-P-FDO outperform all SOTA models with fewer parameters and faster training speed.

\subsubsection{Compatibility analysis.} We further apply FDO to DNN, CN-V2, DCN-V2 and FmFM, which do not require the same field dimensions. Additionally, we compare FDO with assigning fixed dimensions and the Formula method mentioned in Section~\ref{sec:field dimension optimization} to assign various dimensions. Table \ref{tab:compatibility} shows the results. First, the Formula method increases the model parameters and decreases its performance, as it only considers the number of features in each field, disregarding its importance during the training process. In contrast, FDO is a posteriori approach that learns field dimensions by incorporating important information in the trained embedding table. Second, applying FDO to CN-V2 and GCN yields better performance than the base models. Since CN-V2 and GCN mainly focus on explicit bit-level feature crosses, FDO refines the field dimensions by removing unnecessary embedding information. However, applying FDO to DNN, DCN-V2 and GDCN-P, which include a DNN network, leads to a slight performance decrease. Lastly, these results show that we can use FDO to obtain group field dimensions based on the SOTA model and reuse it as the default dimensions for other models. In the last row of Table~\ref{tab:compatibility}, we learn group field dimensions with 95\% information ratio based on the GCN and apply them to the other five models. This approach achieves comparable performance and further reduces the number of parameters compared to using the models themselves to learn the dimensions. 


\begin{table}
\setlength{\abovecaptionskip}{0.2cm}
\setlength{\belowcaptionskip}{-0.2cm}
\centering
\caption{The comparison of different dimension assignment methods. We list the parameter number of all model variants.}
\label{tab:compatibility}
\scalebox{0.84}{
\begin{tabular}{c|cccccc} 
\toprule
Models & CN-V2 & GCN & DNN & DCN-V2 & GDCN-P & FmFM \\ 
\hline
Fixed & \begin{tabular}[c]{@{}c@{}}0.8140\\(18.6M)\end{tabular} & \begin{tabular}[c]{@{}c@{}}0.8154\\(19.7M)\end{tabular} & \begin{tabular}[c]{@{}c@{}}0.8141\\(18.0M)\end{tabular} & \begin{tabular}[c]{@{}c@{}}0.8144\\(19.1M)\end{tabular} & \begin{tabular}[c]{@{}c@{}}0.8161\\(20.3M)\end{tabular} & \begin{tabular}[c]{@{}l@{}}0.8122\\(18.7M)\end{tabular} \\ 
\hline
Formula & \begin{tabular}[c]{@{}c@{}}0.8139\\(20.5M)\end{tabular} & \begin{tabular}[c]{@{}c@{}}0.8154\\(20.7M)\end{tabular} & \begin{tabular}[c]{@{}c@{}}0.8136\\(20.7M)\end{tabular} & \begin{tabular}[c]{@{}c@{}}0.8140\\(20.9M)\end{tabular} & \begin{tabular}[c]{@{}c@{}}0.8157\\(21.1M)\end{tabular} & \begin{tabular}[c]{@{}l@{}}0.8121\\(21.4M)\end{tabular} \\
\hline
\begin{tabular}[c]{@{}c@{}}FDO\\(95\%)\end{tabular} & \begin{tabular}[c]{@{}c@{}}0.8142\\(9.2M)\end{tabular} & \begin{tabular}[c]{@{}c@{}}0.8157\\(7.0M)\end{tabular} & \begin{tabular}[c]{@{}c@{}}0.8140\\(12.2M)\end{tabular} & \begin{tabular}[c]{@{}c@{}}0.8143\\(10.95M)\end{tabular} & \begin{tabular}[c]{@{}c@{}}0.8160\\(7.6M)\end{tabular} & \begin{tabular}[c]{@{}l@{}}0.8122\\(9.4M)\end{tabular} \\ 
\hline
\hline
\begin{tabular}[c]{@{}c@{}}FDO\\(GCN-95\%)\end{tabular} & \begin{tabular}[c]{@{}c@{}}0.8142\\(6.7M)\end{tabular} & \begin{tabular}[c]{@{}c@{}}0.8157\\(7.0M)\end{tabular} & \begin{tabular}[c]{@{}c@{}}0.8138\\(6.9M)\end{tabular} & \begin{tabular}[c]{@{}c@{}}0.8142\\(7.2M)\end{tabular} & \begin{tabular}[c]{@{}c@{}}0.8160\\(7.4M)\end{tabular} & \begin{tabular}[c]{@{}l@{}}0.8122\\(6.5M)\end{tabular} \\ 
\hline
\bottomrule
\end{tabular}
}
\end{table}

\subsubsection{The understanding of condensed field dimensions.} 
\label{sec:relation}
The field dimensions learned through FDO indicate the importance of the corresponding fields. As observed in Figure~\ref{fig:100w_gate values}, we can determine the average importance of each field, such as fields \{\#20, \#23, \#24 \} being important, while fields \{\#11, \#27, \#30\} are considered unimportant. Referring back to Table~\ref{tab:80_90}, fields \{\#20, \#23, \#24\} indeed have longer dimensions after applying FDO with 95\% information ratio. Conversely, fields \{\#11, \#27, \#30\} have shorter dimensions. To further validate this observation, we compute the Pearson Correlation Coefficient between the averaged field importance in the 1st layer and the optimization field dimensions with 95\% ratio. \textit{The correlation coefficient is 0.82 with a p-value less than 1*e-9}, which confirms a significant correlation between the field dimensions and their respective importance. Therefore, we can roughly identify which fields are important directly from the field dimensions learned by FDO. Please note that the field dimension is not always associated with the number of features in the field. For example, fields \{\#16, \#25\} have the maximum feature number, but their field dimensions are short, and their importance is also insignificant. If the Formula method were used to assign field dimensions, fields \{\#16, \#25\} would have the longest dimensions. This comparison further highlights the rationality and superiority of the introduced FDO approach.

\section{Conclusions} 
This paper proposes a novel method GDCN to model both explicit and implicit feature crosses. As the core structure, GCN utilizes the information gate to identify important cross features, avoiding model performance degradation in deeper high-order layers. Importantly, GCN shows great interpretability at both model and instance levels, helping us to understand the model predictions. The proposed FDO approach learns field-specific and condensed dimensions for different fields based on their inherent importance, assisting GCN and GDCN to achieve comparable performance with lighter model size. Extensive experiments verify the remarkable effectiveness and superiority of the GDCN model and FDO approach.

\begin{acks}
This work was supported by the National Natural Science Foundation of China (NSFC) under Grants 62172106 and 61932007.
\end{acks}

\clearpage
\bibliographystyle{ACM-Reference-Format}
\balance
\bibliography{gdcn}

\clearpage
\appendix
\include{appendix}

\end{document}

%% file: appendix.tex
\section{The generalization framework of FDO approach}
\begin{figure}[h]
    \centering
    \includegraphics[width=0.46\textwidth]{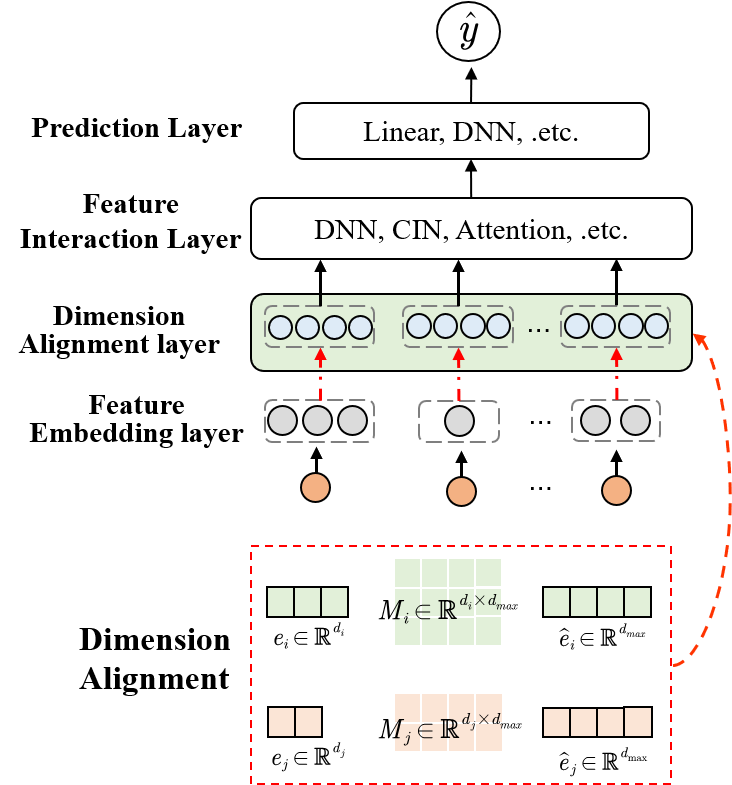}
    \caption{The structure of used generalization framework with dimension alignment layer.}
    \label{fig:framework}
\end{figure}

The proposed GDCN model can receive different dimensions as inputs and use the refined feature dimension information learned by FDO. In addition, methods such as DCN DCN-V2 can also accept different field dimensions as input. However, due to the limitations of the interaction structure, most methods can only accept the same feature representation dimension, such as xDeepFM, FGCNN, AutoInt, and FINT. Therefore, in order to make feature representations of different dimensions that can be used seamlessly in these base models, we use a generalization framework that contains a dimension alignment layer, as shown in Figure \ref{fig:framework}. This method is also used in \cite{qu2022sseds, zhao2021autodim}. Specifically, for each field, we initialize an alignment matrix. For a specific dataset, all features in each field share a dimension alignment matrix, we initialize $F$ alignment matrices $\mathbf{M} = \{\mathbf{M}_1, \mathbf{M}_2,...,\mathbf{M}_F\}$, where $\mathbf{M}_f \in \mathbb{R}^{d_{f}\times d_{max}}$ and $d_{max} = max\{d_1, d_2,...d_F\}\leq D$. $d_{max}$ is the maximum field dimension among all dimensions learned by the FDO approach. With the alignment matrix, we can get the aligned embedding $\mathbf{\hat{e}}_f = \mathbf{e}_f\mathbf{M}_f \in \mathbb{R}^{d_{max}}$. In this process, the dimension alignment layer generates a small number of learnable parameters:
\begin{align}
\mathcal{P}_{align} = \sum_{i=1}^{F}{d_{max}}*d_f.
\end{align}
Take the Criteo dataset as an example, F=39, D=16 and $d_{max} \leq 16$. Hence, the added parameters $\mathcal{P}_{align}$ is less than 9984. Compared to the feature representation parameters that can be reduced, these parameters are negligible. After dimension alignment, most feature-based interactions can be performed. Based on the dimension alignment framework, other CTR models can also accept the refined dimensions learned by the FDO approach to reduce model parameters seamlessly and also improve training efficiency. Furthermore, subsequent experiments verify that applying the framework to basic model can improve their prediction performance.

\section{Added Experiments}
In addition to the experimental results that already given in Section \ref{sec:exp_all}, we further share several relevant and necessary experiments and corresponding results.

\begin{figure}[t]
\centering
\subfloat[Field \#2]{
\begin{minipage}[t]{0.49\linewidth}
\centering
\includegraphics[width=1.0\textwidth]{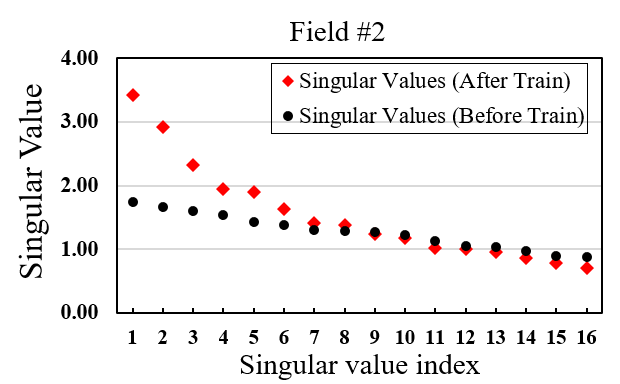}
\end{minipage}
}
\subfloat[Field \#2, $d_2=13$]{
\begin{minipage}[t]{0.49\linewidth}
\centering
\includegraphics[width=1.0\textwidth]{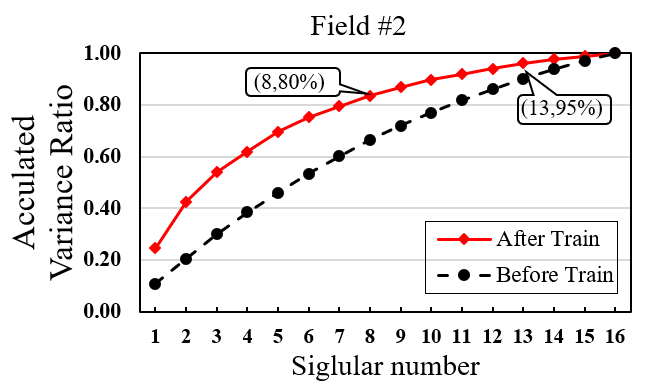}
\end{minipage}
}

\subfloat[Field \#6]{
\begin{minipage}[t]{0.49\linewidth}
\centering
\includegraphics[width=1.0\textwidth]{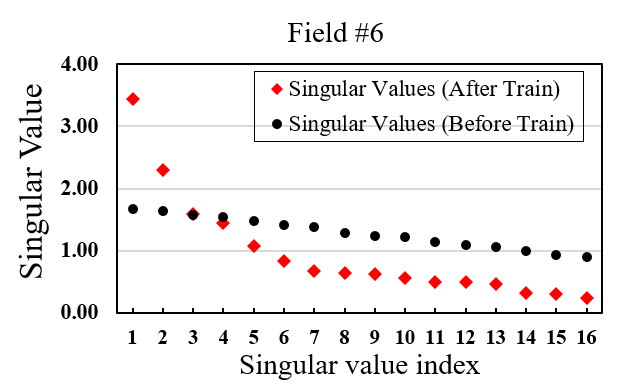}
\end{minipage}
}
\subfloat[Field \#6, $d_6=9$]{
\begin{minipage}[t]{0.49\linewidth}
\centering
\includegraphics[width=1.0\textwidth]{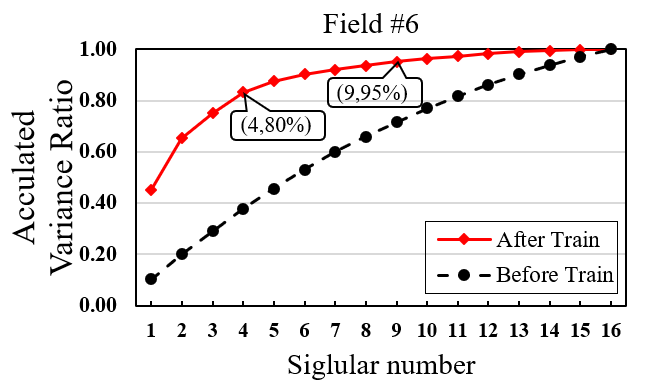}
\end{minipage}
}

\subfloat[Field \#16]{
\begin{minipage}[t]{0.49\linewidth}
\centering
\includegraphics[width=1.0\textwidth]{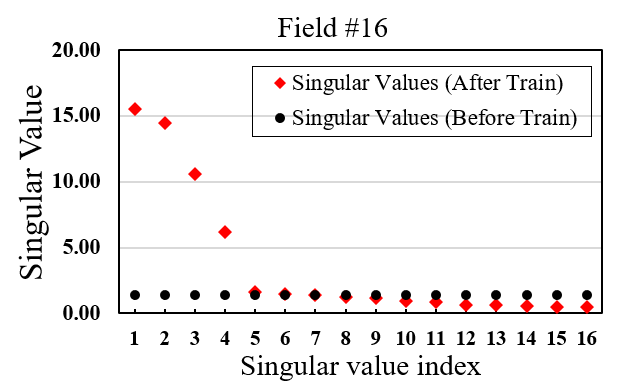}
\end{minipage}
}
\subfloat[Field \#16, $d_{16}=4$]{
\begin{minipage}[t]{0.49\linewidth}
\centering
\includegraphics[width=1.0\textwidth]{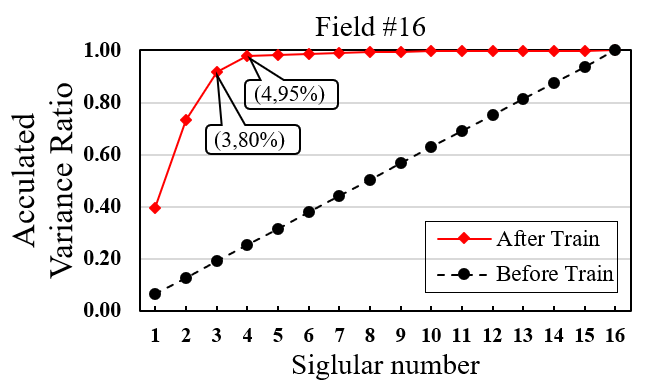}
\end{minipage}
}

\centering
\caption{Schematic diagram of the FDO method for calculating feature dimensions based on the kept information ratio.}
\label{fig:fdo_understanding}
\end{figure}

\subsection{Dimension Understanding}

Figure \ref{fig:fdo_understanding} illustrates the process of the FDO method to select the appropriate dimension based on the kept information ratio. We select three representative fields (i.e., fields {\#2, \#6 and \#16}) for further analysis. Subplots \ref{fig:fdo_understanding} (a), (b), and (c) list the calculated singular values (sorted from largest to smallest) based on the feature embedding matrix of corresponding fields before and after model training. In those subplots, the black points indicate the calculated singular values based on the initial feature embedding matrix, and the red points indicate the calculated singular values based on the trained feature embedding matrix. The corresponding subplots (b), (e), and (f) plot the ratio of cumulative information based on calculated singular values, where the horizontal coordinate indicates the number of singular values, ranging from 1 to 16. Also, the figure lists the minimum number of singular values required to achieve the specified information ratio. For example, (13, 95\%) in subplot (b) indicates that a minimum of 13 singular values are required for the sum of their information variance to exceed 95\% of the total information. Therefore, for field \#2, the feature dimension that should be assigned while retaining 95\% information ratio is 13. Furthermore, from table \ref{tab:80_90}, we know the feature number contained in the fields \{\#2, \#6, \#16\} are 101, 118, and 193,949, respectively. Based on the above information, the following observations can be observed:

Firstly, the singular values learned based on the initial feature embedding matrix do not provide valid information. As can be seen from subplots (a), (c) and (e), the difference in singular values in the same subplot is small at this point. Especially when the number of features contained in the field is large (subplot (e)), all the singular values are basically equal. From subplots (b), (d) and (f), it can be observed that the proportion of cumulative information calculated based on these singular values increases almost uniformly.

Secondly, the training-based feature embedding matrix can provide significant instruction for dimension selection. After model training, the importance of the feature field is implicitly reflected in the feature embedding matrix and can be expressed explicitly through the informative magnitude of the singular values. While keeping 95\% information ratio, the calculated feature dimensions for the three feature fields are 13, 9 and 4, respectively. Comparing the red scatter plots in subfigures (a), (c), and (e), the smaller dimension required for the feature field corresponds to a larger singular value, and the gap between the maximum and minimum singular values will be larger. Larger singular values take up a larger proportion of the information, so subfigure (f) needs only 4 singular values to obtain 95\% information ratio. For field \#16, other dimensions contain very little information but occupy a large number of model parameters and even inversely affect the predictive performance.

\subsection{Effectiveness Analysis}
\begin{table}[b]
\centering
\caption{The effectiveness of GCN with FDO approach on the Malware dataset.} 
\label{tab:variance_malware}
\scalebox{0.90}{
\begin{tabular}{c|cc|cc|cc}
\hline
\hline
\multirow{1}{*}{\begin{tabular}[c]
{@{}c@{}}Ratio\end{tabular}} & \multirow{1}{*}{$\overline{D}$} & \multirow{1}{*}{$\overline{K}$} & \multicolumn{1}{c}{\multirow{1}{*}{\begin{tabular}[c]{@{}c@{}}\#Params\end{tabular}}}&\multicolumn{1}{c|}{Proportion} & \multicolumn{1}{c}{AUC} &\multicolumn{1}{c}{Logloss}  \\ 
\hline
Full (16) & 16  & 16  & 25.70M&100\%&0.7445&0.5908\\
98\%& 4.88& 3.64& 5.29M&20.6\%&0.7447	&0.5904\\
95\%& 3.57& 2.94& 3.82M&14.9\%&0.7445	&0.5907\\
90\%& 2.40& 2.38& 2.56M&10.0\%&0.7445	&0.5908\\
80\%& 1.17& 1.79& 1.46M&5.7\%&0.7442	&0.5910\\
70\%& 1.11& 1.52& 1.18M&4.6\%&0.7429	&0.5920\\
60\%& 1.06& 1.31& 1.10M&4.3\%&0.7419	&0.5929\\
50\%& 1.01& 1.12& 1.04M&4.0\%&0.7396	&0.5947\\
\hline
\hline
Full (8)& 8  & 8  & 10.33M&40.2\%&0.7442	&0.5910\\
\hline\hline
\end{tabular}}
\end{table}
Table \ref{tab:variance_malware} is the supplement experiment of Section \ref{exp:dimension optimization}, which show the experiment results of the Malware dataset with different information ratio. 

Similar to the Criteo dataset, when maintaining 80\% information ratio, the model's performance is on par with the full model. For the Malware dataset, the FDO method is more effective in reducing the model parameters. Specifically, when the kept information ratio is between 80\% to 98\%, the corresponding model parameters are only 1.46M and 5.29M, only occupying 5.7\% to 20.6\% of the full model. Meanwhile, compared to basic stacked and parallel models (as shown in Table \ref{tab:all} and \ref{tab:parallel}), GDCN-FDO (with 80\% information ratio) enables to achieve better performance with only 1.46M model parameters.

\begin{figure}[h]
    \centering
    \includegraphics[width=0.45\textwidth]{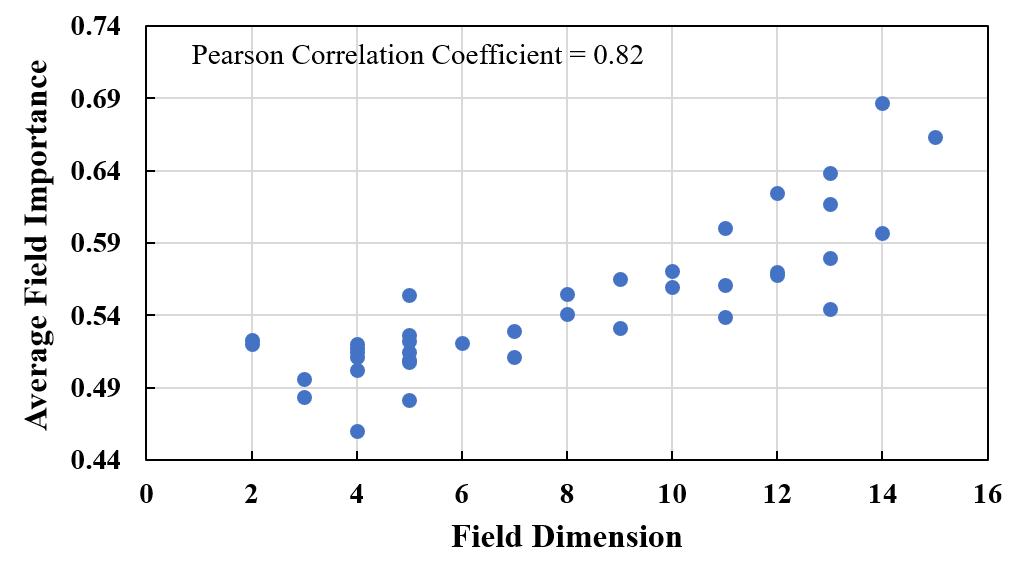}
    \caption{The visualization of the relationship between averaged field importance and field dimension.}
    \label{fig:relation}
\end{figure}

\subsection{Visualization of Relationship}

In section \ref{sec:relation}, we confirm the significant positive correlation between the field dimensions and their respective importance. In Figure \ref{fig:relation}, we visualize the positive correlation between field dimensions and field importance through a scatter plot. It is clear to see that when the feature representation dimension is larger, the average importance of the feature is also more significant. In contrast, the dimension of a field is not directly related to the number of features it contains. Specifically, the Pearson Correlation Coefficient is -0.23 for these two groups of data. For example, fields \{\#16, \#25\} contain the most feature number but exhibit smaller dimensions. If an empirical formula is used to calculate feature dimensions, fields \{\#16, \#25\} will be assigned the longest dimension.

\subsection{The Compatibility of the Generalization Framework}

\begin{table}[bh]
\centering
\caption{The Compatibility of the Generalization Framework}
\label{tab:compatibility_fra}
\scalebox{0.82}{
\begin{tabular}{c|ccc|ccc} 
\hline
\hline
Datasets & \multicolumn{3}{c|}{\textbf{Criteo}} & \multicolumn{3}{c}{\textbf{Malware}} \\ 
\hline
Models & AUC $\uparrow$ & Logloss$\downarrow$ & \#Params & AUC$\uparrow$ & Logloss $\downarrow$ & \#Params \\ 
\hline
FM & 0.8085 & 0.4433 & 18.48M & 0.7363&0.5982& 16.60M \\
FM-FDO & 0.8099 & 0.4418 & 7.52M & 0.7368 & 0.5973 & 4.46M \\ 
$\Delta$&\textbf{0.0014} &\textbf{-0.0015} &\textbf{40.7\%} &\textbf{0.0005} &\textbf{-0.0009} &\textbf{26.9\%}\\
\hline
IPNN & 0.8128 &0.4390 & 18.26M & 0.7433 &0.5918 & 17.76M \\
IPNN-FDO & 0.8130 & 0.4388 & 7.29M & 0.7436 & 0.5916 & 5.43M \\ 
$\Delta$&\textbf{0.0002} &\textbf{-0.0003} &\textbf{39.9\%} &\textbf{0.0003 }&\textbf{-0.0002} &\textbf{30.6\%}\\
\hline
FINT & 0.8128 &0.4390 & 19.05M & 0.7424 &0.5924 & 17.46M \\
FINT-FDO & 0.8132 & 0.4387 & 8.84M & 0.7430 & 0.5920 & 5.13M \\ 
$\Delta$&\textbf{0.0004} &\textbf{-0.0003} &\textbf{46.4\%} &\textbf{0.0006} &\textbf{-0.0004} &\textbf{29.4\%}\\
\hline
AFN & 0.8116 &0.4401 &17.78M & 0.7427 &0.5924 & 16.98M \\
AFN-FDO & 0.8132 & 0.4386 & 7.91M & 0.7429 &0.5921 & 4.83M \\ 
$\Delta$&\textbf{0.0016} &\textbf{-0.0015} &\textbf{44.5\%} &\textbf{0.0002} &\textbf{-0.0003} &\textbf{28.4\%}\\
\hline
CIN & 0.8121 &0.4398 &17.93M & 0.7424 &0.5928 & 17.09M \\
CIN-FDO &0.8129 &0.4388&6.98M&0.7430 &0.5921 & 4.95M\\
$\Delta$ &\textbf{0.0008} &\textbf{-0.0010} &\textbf{38.9\%} & \textbf{0.0006} & \textbf{-0.0007} &\textbf{29.0\%}\\
\hline
\hline
\end{tabular}}
\end{table}

Based on the proposed generalization framework, most deep CTR models can receive different feature dimensions as inputs. Table \ref{tab:compatibility_fra} lists the improvement in prediction performance and model parameters before and after applying the framework. The adopted field dimension is learned by the FDO approach with 95\%  kept information ratio.  

Firstly, after applying the generalization framework, the prediction accuracy of the five base models significantly improves. For example, the FM and AFN in the Criteo dataset have an AUC boost of 0.0014 and 0.0016. Other models can also obtain a slight effect boost, with AUC boosts ranging from 0.0002 to 0.0006 and Logloss optimization ranging from 0.0003 to 0.0009.

Secondly, using different feature dimensions can significantly reduce the overall parameters of the underlying CTR prediction model. For the Criteo dataset, the model parameters are reduced to 39.9\% to 46.4\%; for the Malware dataset, the model parameters are reduced to 26.9\% to 30.6\%. More importantly, the performance of these models is maintained. 

In summary, applying a dimensional alignment layer allows most CTR prediction models to receive embedded features of different dimensions, which reduces the overall parameters of the model while improving its predictive power.